\documentclass[journal]{IEEEtran}

\IEEEoverridecommandlockouts

\usepackage{cite}
\usepackage{amsmath, amssymb, amsfonts, bm}
\usepackage{amsthm}
\usepackage{algorithmic}
\usepackage[ruled, linesnumbered]{algorithm2e}
\usepackage{setspace}
\usepackage{indentfirst}
\usepackage{graphicx}
\usepackage{epstopdf}
\usepackage{textcomp}
\usepackage{xcolor}
\usepackage{subfigure}
\usepackage{mathrsfs}
\usepackage{verbatim}
\usepackage{url}
\usepackage{multirow}
\usepackage{multicol}
\usepackage{booktabs}
\usepackage{stfloats}
\usepackage{colortbl}
\usepackage{makecell}
\usepackage{tabularx}
\usepackage{mathtools}
\usepackage{pifont}
\usepackage{array}
\usepackage{float}
\usepackage{threeparttable}
\usepackage{cases}
\usepackage{bbm}

\usepackage{tikz}
\usetikzlibrary{arrows, shapes, chains}

\usepackage{color}
\definecolor{color1}{RGB}{128, 0, 0}
\definecolor{color2}{RGB}{0, 64, 0}
\definecolor{color3}{RGB}{0, 0, 128}
\usepackage[colorlinks, linkcolor=color1, anchorcolor=blue, citecolor=color3, urlcolor = color2]{hyperref}

\newtheorem{theorem}{Theorem}

\newtheorem{definition}{Definition}

\allowdisplaybreaks[4]

\begin{document}

\title{Over-the-Air Computation for 6G: Foundations, Technologies, and Applications}

\author{
	\IEEEauthorblockN{Zhibin Wang\IEEEauthorrefmark{1}, \IEEEmembership{Graduate Student Member, IEEE}, Yapeng Zhao\IEEEauthorrefmark{1}, Yong Zhou, \IEEEmembership{Senior Member, IEEE}, \\
		Yuanming Shi, \IEEEmembership{Senior Member, IEEE},	Chunxiao Jiang, \IEEEmembership{Fellow, IEEE}, and Khaled B. Letaief, \IEEEmembership{Fellow, IEEE}}
	
	\thanks{\IEEEauthorrefmark{1}Equally contributed.}
	\thanks{
		Zhibin Wang, Yong Zhou, and Yuanming Shi are with the School of Information Science and Technology, ShanghaiTech University, Shanghai 201210, China (e-mail: \{wangzhb, zhouyong, shiym\}@shanghaitech.edu.cn).
	}
	\thanks{
		Yapeng Zhao is with the State Key Laboratory of Internet of Things for Smart City, University of Macau, Macao 999078, China (e-mail: yc17435@connect.um.edu.mo).
	}
	\thanks{
		Chunxiao Jiang is with the Tsinghua Space Center and the Beijing National Research Center for Information Science and Technology, Tsinghua University, Beijing 100084, China (e-mail: jchx@tsinghua.edu.cn).
	}
	\thanks{
		Khaled B. Letaief is with the Department of Electronic and Computer Engineering, The Hong Kong University of Science and Technology, Hong Kong (e-mail: eekhaled@ust.hk).
	}
}

\maketitle

\begin{abstract}

The rapid advancement of artificial intelligence technologies has given rise to diversified intelligent services, which place unprecedented demands on massive connectivity and gigantic data aggregation.
However, the scarce radio resources and stringent latency requirement make it challenging to meet these demands.
To tackle these challenges, \textit{over-the-air computation} (AirComp) emerges as a potential technology.
Specifically, AirComp seamlessly integrates the communication and computation procedures through the superposition property of multiple-access channels, which yields a revolutionary multiple-access paradigm shift from ``compute-after-communicate'' to ``compute-when-communicate''.
By this means, AirComp enables spectral-efficient and low-latency wireless data aggregation by allowing multiple devices to occupy the same channel for transmission.
In this paper, we aim to present the recent advancement of AirComp in terms of foundations, technologies, and applications.
The mathematical form and communication design are introduced as the foundations of AirComp, and the critical issues of AirComp over different network architectures are then discussed along with the review of existing literature.
The technologies employed for the analysis and optimization on AirComp are reviewed from the information theory and signal processing perspectives.
Moreover, we present the existing studies that tackle the practical implementation issues in AirComp systems, and elaborate the applications of AirComp in Internet of Things and edge intelligent networks.
Finally, potential research directions are highlighted to motivate the future development of AirComp.
	
\end{abstract}

\begin{IEEEkeywords}
	
Over-the-air computation, integrated communication and computation, federated learning.
	
\end{IEEEkeywords}

\section{Introduction} \label{sec:intro}

While the \textit{fifth-generation} (5G) wireless networks are being deployed worldwide, the preliminary outlook for the next generation of networks, referred to as the \textit{sixth-generation} (6G) wireless communication systems, has been initiated by both academia and industry.
Since 6G is still in its infancy, a number of envisioned applications, system requirements, and innovative techniques are spotlighted and discussed in a growing body of works \cite{letaief2019roadmap, saad2020vision, yang2020artificial, tataria20216g, letaief2022edge, shi2023task}.
A prevailing view among these works is that 6G will give rise to numerous intelligent services with the advancement of emerging \textit{artificial intelligence} (AI) techniques, especially deep learning and reinforcement learning.
Meanwhile, massive connectivity and tremendous information exchange are imperative to fulfill the needs of data analysis and model training for intelligent services, which, however, poses significant challenges to current resource-constrained wireless communication systems.
Accordingly, efficient and scalable multiple-access protocols are required to enable ubiquitous connections for rapidly growing devices.

In retrospect, the increasing demand for wireless connections has stimulated the rapid development of multiple-access strategies over the previous decades.
The existing multiple-access strategies can generally be divided into two categories, i.e., orthogonal and non-orthogonal schemes.
Specifically, \textit{orthogonal multiple access} (OMA) schemes entitle each of the devices to have an exclusive occupation of specifically allocated resources, which leads to interference-free transmission and results in a simple transceiver design.
Depending on the allocated orthogonal domains, several OMA schemes have been extensively adopted in current wireless systems, including frequency-division multiple access, time-division multiple access, code-division multiple access \cite{gilhousen1991on}, and orthogonal frequency-division multiple access \cite{yin2006ofdma}.
Since the available resources are divided into orthogonal blocks, OMA strictly limits the total number of simultaneously scheduled devices under limited radio resources.
To circumvent such a restriction, \textit{non-orthogonal multiple access} (NOMA) schemes become potential candidates for future massive access networks \cite{liu2022evolution}, such as power-domain NOMA \cite{dai2015noma, zhou2018dynamic, zhou2018stable, zhou2018coverage, zhou2022performance}, code-domain NOMA \cite{dai2018survey}, space-division multiple access \cite{agiwal2016next}, and rate-splitting multiple access \cite{clerckx2016rate, mao2022ratesplitting}, in which one resource block can be shared by multiple devices for simultaneous transmissions at the cost of introducing co-channel interference.
Several techniques have been proposed for tackling the interference caused by overlapped resource allocation, such as \textit{successive interference cancellation} (SIC), message passing, and multi-antenna beamforming.
Assisted by these interference cancellation methods, NOMA is capable of yielding significant gains in massive connectivity, spectral efficiency, and communication latency \cite{dai2018survey, liu2022evolution}.

The aforementioned multiple-access strategies are primarily designed according to Shannon-Hartley theorem \cite{shannon1948mathematical}, with an objective to achieve maximum reliable data transmission rate.
As these protocols disregard the concrete tasks for which the data is used, they can be deemed as task-agnostic multiple access, where the transmission and the utilization of data are treated as two separate and independent parts \cite{zhu2021overtheair}.
However, such an isolated design inevitably prevents the inherent correlation between the transmitted data and subsequent tasks from being exploited to further promote the communication efficiency, especially in the scenarios that require immediate decision makings in response to the delivered data, such as security alerts in environmental sensing and obstacle avoidance in autonomous driving.
Therefore, how to enable task-oriented multiple access becomes a potential research direction for future wireless networks.
Fortunately, one observation can be obtained that supporting the tasks of intelligent services generally demands the function computation on dispersed data for the final decision, such as computing the average temperature acquired by distributed sensors for fire warning and computing the average reflection delay of signals radiated by multiple sensors for obstacle location estimation.
Accordingly, the function computation can be perceived as establishing a connection between the data and its service task, hence stimulating the exploration of task-oriented multiple access by integrating the communication and computation.
Meanwhile, low communication latency is an essential requirement in intelligent services to guarantee the timeliness of corresponding applications, which further motivates the low-latency multiple-access design.
With the above motives in mind, \textit{over-the-air computation} (AirComp) becomes a promising candidate to enable efficient \textit{wireless data aggregation} (WDA) with the fusion of communication and computation \cite{zhu2021overtheair}.

The kernel of AirComp is to integrate the communication and computation via the waveform superposition property of \textit{multiple-access channels} (MACs) \cite{nazer2007computation}, where the concurrent signals can naturally execute an additive operation over the same radio channel.
Unlike conventional multiple-access schemes considering co-channel interference as a disruptor of wireless transmission, AirComp treats the interference as a contributor to the function computation and forgoes the interference cancellation for decoding each of the data.
Therefore, in terms of the service task of computing functions, AirComp is capable of achieving higher spectral efficiency by letting all devices share the full bandwidth and lower communication latency by eliminating the necessity to demodulate the data one by one as compared with conventional task-agnostic multiple access schemes.
It is worth noting that the similar idea to exploiting waveform superposition has been previously harnessed by \textit{physical-layer network coding} (PNC) to improve the network throughput \cite{zhang2006hot, liew2013physical, chen2020physical}, which enables combined simultaneously transferred messages to be relayed over wireless networks.
Even though, the receiver in PNC-aided networks still needs to extract its required information from the overlapped signals, which differs from the purpose of AirComp to estimate a function of the received signal.
Besides, the idea of data fusion with superimposed signals has also been adopted to tackle \textit{chief executive officer} (CEO) problems \cite{berger1996ceo, gastpar2003source, gastar2008uncoded}, which aims at estimating a remote common source from multiple noisy observations in wireless sensor networks.
In contrast to the CEO problem that intends to realize a single-source estimation, AirComp desires to achieve a function computation of diverse data from multiple sources, which is fundamentally different from the CEO problem in terms of the problem objective and the system architecture.

The study on AirComp is initiated by the seminal work \cite{nazer2007computation}, which pioneers the use of computation coding for reliable function computation over MACs.
Different from conventional information-theoretic coding strategies, computation coding does not force the transmit signals to be represented in digital bits \cite{nazer2007computation}, leading to the subsequent development of AirComp being divided into two branches, i.e., coded/digital and uncoded/analog AirComp.
In the meantime, two novel performance metrics, namely computation rate and computation accuracy, are introduced to measure the reliability of the function reconstruction, which respectively evaluate the maximum number of functions calculated per channel use and the distortion between the estimated and the desired function values \cite{goldenbaum2014on, goldenbaum2015nomographic}.
Under the guidance of these two metrics, several analyses and transceiver design for AirComp over multifarious network architectures have been developed recently.
Furthermore, the unprecedented growth of \textit{Internet of Things} (IoT) and \textit{edge intelligent} (EI) networks prompts a growing enthusiasm for AirComp, motivating the researchers to further explore its implementation in emerging applications and potential in driving the evolution of wireless networks.
As a consequence, the research on AirComp has become active and numerous remarkable studies in this field have also proliferated recently.

\subsection{Motivations and Contributions}

\begin{table*}[t]
	\centering
	\caption{Comparison Between Our Paper and Existing Reviews on AirComp}
	\label{tab:comparison}
	\scriptsize
	\renewcommand{\arraystretch}{1.5}
	\begin{tabular}{!{\vrule width1pt}c!{\vrule width1pt}c|c|m{0.25\linewidth}|m{0.45\linewidth}!{\vrule width1pt}}
		\Xhline{1pt}
		\rowcolor[HTML]{e8e8e8}
		\bf{Reference} & \bf{Year} & \bf{Type} & \multicolumn{1}{c|}{\bf{Similarities}} & \multicolumn{1}{c!{\vrule width1pt}}{\bf{Differences}} \\ 
		\Xhline{1pt}
		\cite{zhu2021overtheair} & 2021 & Overview & Similar to our survey, this overview introduces basic principles, enabling techniques, and applications of AirComp through several typical studies. & This overview provides a concise overview to facilitate the understanding of AirComp, while our survey delivers a more detailed and systematic review to encompass the large body of existing research on AirComp. \\ 
		\rowcolor[HTML]{f4f4f2}
		\cite{altun2022magic} & 2022 & Survey & Similar to our survey, this survey highlights the research on exploring the superposition property of MACs for integrating communication and computation with AirComp. & This survey investigates the techniques that harness the superposition property of MACs for various use cases, e.g., compute-and-forward, type-based multiple access, and computation over multiple access. In contrast, our paper focuses on AirComp that is leveraged for computation over multiple access in a more detailed and systematic way. \\
		\cite{chen2021distributed} & 2021 & Survey & Similar to our survey, this survey indicates the advantage of AirComp for enabling communication-efficient model aggregation in distributed learning scenarios. & This survey presents the review on the communication technologies for efficiently embedding distributed learning in wireless networks, where AirComp is referred as one of advanced techniques for efficient model aggregation. 
		In contrast, our paper not only discusses several applications of AirComp in both IoT and EI networks, including the model aggregation for distributed learning, but also presents detailed and systematic review on advanced techniques for AirComp. \\
		\rowcolor[HTML]{f4f4f2}
		\cite{sahin2023surveyaircomp} & 2023 & Survey & Similar to our survey, this survey presents a systematic and comprehensive review on AirComp. & This survey categorizes the existing AirComp techniques according to physical layer communication process, while our survey discusses the literature on the basis of various network architectures and analyses/algorithms for transceiver design. \\
		\Xhline{1pt}
	\end{tabular}
\end{table*}

Driven by the rapid advancement of AirComp, several discussions and overviews on AirComp have been made in recent works.
For instance, a concise overview is presented in \cite{zhu2021overtheair} to briefly introduce the background and applications of AirComp, but lacks a comprehensive presentation and categorization of the large body of recent available research on AirComp.
The authors in \cite{altun2022magic} review the technologies that explore the benefits of signal superposition for enhancing the communication efficiency in multiple access networks, where AirComp as one of the presented techniques is sketched out in the article.
AirComp is also reviewed in \cite{chen2021distributed} for efficient wireless model aggregation in distributed learning systems with a dearth of extensive discussions on AirComp.
Therefore, to capture the recent progress and facilitate further development and interdisciplinary research on AirComp, in this paper, we aim to provide a systematic review of existing studies on AirComp, while discussing the existing challenges and envisioning future research directions.
It is noteworthy that a parallel work to ours is conducted in \cite{sahin2023surveyaircomp} to survey how to reliably and efficiently compute a function via AirComp from the physical layer communication perspective, which can complement this work in terms of diverse categorizations.
The comparison between our paper and existing reviews on AirComp is presented in Table \ref{tab:comparison}.
The main contributions of this paper are summarized as follows:
\begin{itemize}
	\item
	Foundations of AirComp are introduced in terms of the computation design and communication design, which establish the theoretical bases of AirComp and demonstrate the feasibility of AirComp in wireless communications.
	
	\item
	Existing works on AirComp over different network architectures are comprehensively reviewed and the major issues in each network architecture are discussed to highlight future research opportunities.
	
	\item
	Analyses and optimization for AirComp are outlined from the information theory and signal processing perspectives, where the system performance and transceiver design are presented according to different network settings.
	
	\item
	Practical implementation issues of AirComp are discussed along with the solutions developed by existing studies, which illustrate the challenges when moving from theoretical research to practical applications.
	
	\item
	Advantages of AirComp when applied to IoT and EI networks are introduced, and future research directions are presented to further promote the development of AirComp in future wireless networks.
	
\end{itemize}

\textit{Organization:}
The structure of this paper is organized as follows.
Section \ref{sec:fundamentals} introduces the basics of AirComp, including the computation design and communication design.
Section \ref{sec:network_architecture} presents a literature review on AirComp over different network architectures.
The analyses and transceiver design from information theory and signal processing perspectives for AirComp are highlighted in Section \ref{sec:information_theory_tools} and Section \ref{sec:signal_processing_tools}, respectively.
Practical implementation issues are discussed in Section \ref{sec:practical_implementation}, and the applications of AirComp in IoT and EI networks are presented in Section \ref{sec:applications}.
Moreover, Section \ref{sec:future} illustrates the potential research topics of AirComp.
Finally, this paper is concluded in Section \ref{sec:conclusion}.

\textit{Notations:}
$\mathbb{R}$ and $\mathbb{C}$ denote real and complex domains, respectively.
$\mathbb{Z}_p = \{0, 1, \dots, p - 1\}$ denotes the integers modulo $p$, where $p$ is a prime number.
$\mathbb{P}[\cdot]$ and $\mathbb{E}[\cdot]$ represent the statistical probability and expectation, respectively.


\section{Basics of AirComp} \label{sec:fundamentals}

This section presents the basics of AirComp from the perspectives of computation design and communication design, which reveal the unique features and benefits of AirComp.

\subsection{Computation Design}

The main purpose of AirComp is to compute a class of so-called nomographic functions at the \textit{fusion center} (FC), e.g., \textit{base station} (BS) and \textit{access point} (AP), by merging the data concurrently transmitted from dispersed devices, where the nomographic function is defined as follows.

\begin{definition}[Nomographic Function \cite{buck1979approximate, goldenbaum2015nomographic}] \label{def:nomo_func}
	A function $f(\cdot): \mathbb{R}^K \rightarrow \mathbb{R}$ with $K \ge 2$ that can be represented in the form of $f(\bm{x}) = \psi\left(\sum_{k = 1}^K \varphi_k(x_k)\right)$ is called a nomographic function if there exist functions $\psi(\cdot): \mathbb{R} \rightarrow \mathbb{R}$ and $\varphi_k(\cdot): \mathbb{R} \rightarrow \mathbb{R}$, $\forall \, k \in \{1, 2, \dots, K\}$, where $\bm{x} = [x_1, x_2, \dots, x_K]^{\sf T}$.
\end{definition}

According to Definition \ref{def:nomo_func}, the data, $\{x_k\}$, generated at each of the devices goes through three procedures to enable the computation of a specific nomographic function at the receiver \cite{zhu2019mimo}:
1) data $x_k$ is pre-processed at device $k$ via function $\varphi_k(\cdot)$; 
2) pre-processed data $\{\varphi_k(x_k)\}$ are summed together via the waveform superposition property of MACs;
and 3) the aggregated data is post-processed at the FC via function $\psi(\cdot)$.
Although the waveform superposition is a summation procedure, applying particularly designed pre- and post-processing functions enables AirComp to go beyond the computation of simple linear functions as averaging and summation \cite{buck1979approximate, abari2016over}, and the examples of computable nomographic functions are presented in Table \ref{tab:nomo_func}.
Besides, the outcomes given in \cite{kolmogorov1957representation, sternfeld1985dimension} reveal that $2 K + 1$ nomographic functions are sufficient to represent any continuous function of $K$ variables, which can be summarized as follows.
\begin{theorem}[Nomographic Representation \cite{kolmogorov1957representation, sternfeld1985dimension, goldenbaum2015nomographic}] \label{theorem:cont_func_nomo_func}
	Every continuous function $\tilde{f}(\cdot): \mathbb{R}^K \rightarrow \mathbb{R}$ can be represented as the summation of no more than $2 K + 1$ nomographic functions, i.e., $\tilde{f}(\bm{x}) = \sum_{j = 1}^{2 K + 1} f_j(\bm{x}) = \sum_{j = 1}^{2 K + 1} \psi_j \left(\sum_{k = 1}^K \varphi_{j, k}(x_k)\right)$, where both pre-processing functions $\varphi_{j, k}(\cdot)$ and post-processing functions $\psi_j(\cdot)$ are continuous, $\forall \, j \in \{1, 2, \dots, 2K + 1\}$, $\forall \, k \in \{1, 2, \dots, K\}$. 
\end{theorem}
Theorem \ref{theorem:cont_func_nomo_func} demonstrates that any continuous function with $K$ variables is computable via AirComp by utilizing at most $2 K + 1$ channel uses with appropriately designed continuous pre- and post-processing functions.
This consolidates the mathematical foundation of AirComp and makes it applicable to compute various desired functions in different practical scenarios.
The above functions can be applied to the complex domain to match the system design of wireless networks.

\begin{table}[t]
	\centering
	\caption{Examples of Computable Nomographic Functions via AirComp}
	\label{tab:nomo_func}
	\scriptsize
	\renewcommand{\arraystretch}{1.5}
	\resizebox{\linewidth}{!}{
		\begin{tabular}{!{\vrule width1pt}c!{\vrule width1pt}c|c|c!{\vrule width1pt}}
			\Xhline{1pt}
			\rowcolor[HTML]{e8e8e8}
			\bf{Function} & \bf{$\varphi_k(x_k)$} & \bf{$\psi(y)$} & \bf{$f(x_1, x_2, \dots, x_K)$} \\ 
			\Xhline{1pt}
			Arithmetic mean & $x_k$ & $y / K$ & $(1 / K) \sum_{k = 1}^K x_k$  \\ 
			\rowcolor[HTML]{f4f4f2}
			Weighted sum & $\alpha_k x_k$ & $y / (\sum_{k = 1}^K \alpha_k)$ & $\sum_{k = 1}^K \alpha_k x_k / (\sum_{k = 1}^K \alpha_k)$  \\
			Geometric mean & $\ln(x_k)$ & $\exp(y / K)$ & $(\prod_{k = 1}^K x_k)^{1 / K}$ \\
			\rowcolor[HTML]{f4f4f2}
			Euclidean norm & $x_k^2$ & $\sqrt{y}$ & $\sqrt{\sum_{k = 1}^K x_k^2}$ \\ 
			Polynomial & $\alpha_k x_k^{\beta_k}$ & $y$ & $\sum_{k = 1}^K \alpha_k x_k^{\beta_k}$ \\ 
			\Xhline{1pt} 
		\end{tabular}
	}
\end{table}

\subsection{Communication Design} \label{subsec:comm_design}

\begin{figure}[t]
	\centering
	\subfigure[``Compute-after-communicate'' strategy]{
		\centering
		\label{fig:comp_after_comm}
		\includegraphics[scale=0.55]{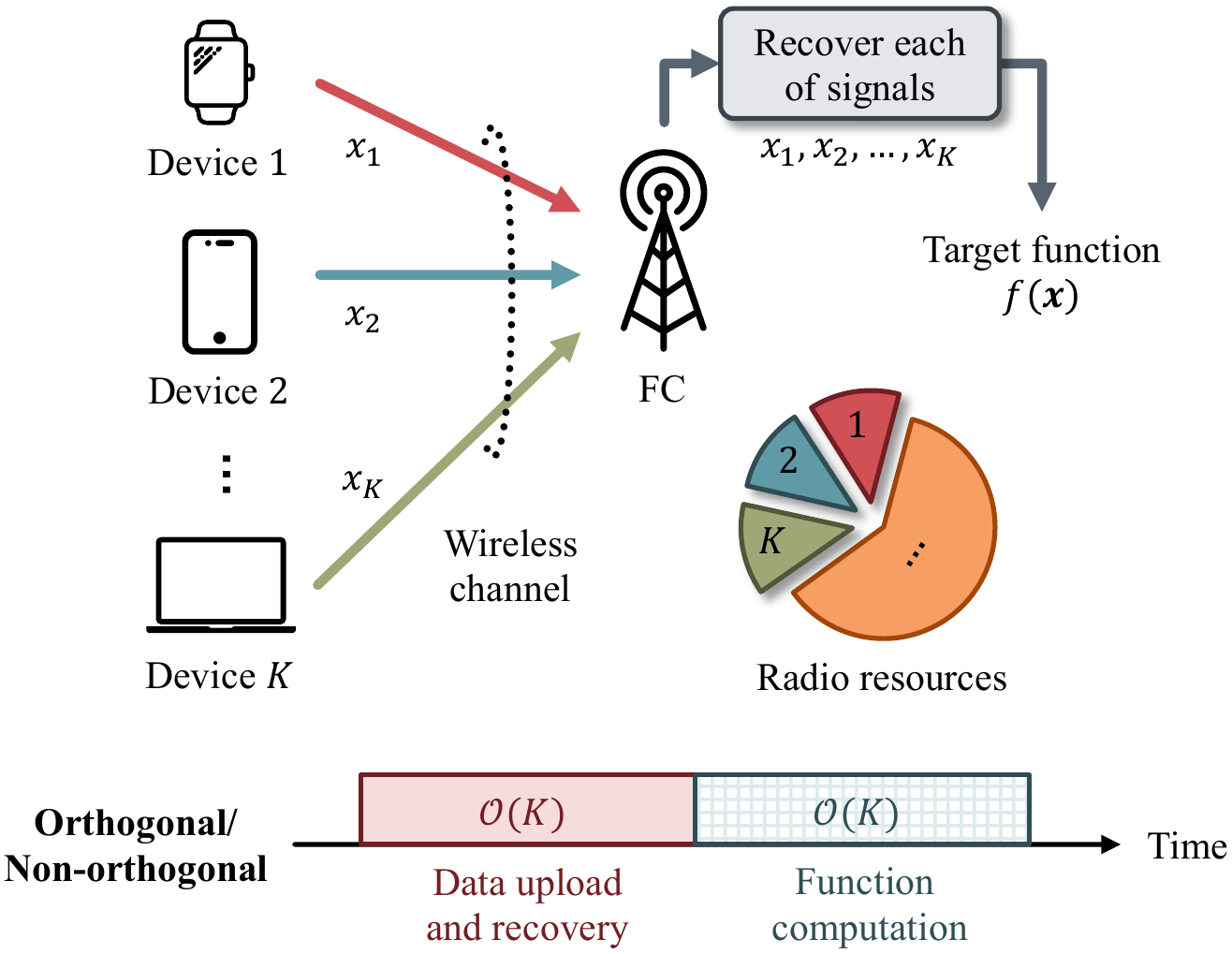}
	}
	\subfigure[``Compute-when-communicate'' strategy]{
		\centering
		\label{fig:comp_when_comm}
		\includegraphics[scale=0.55]{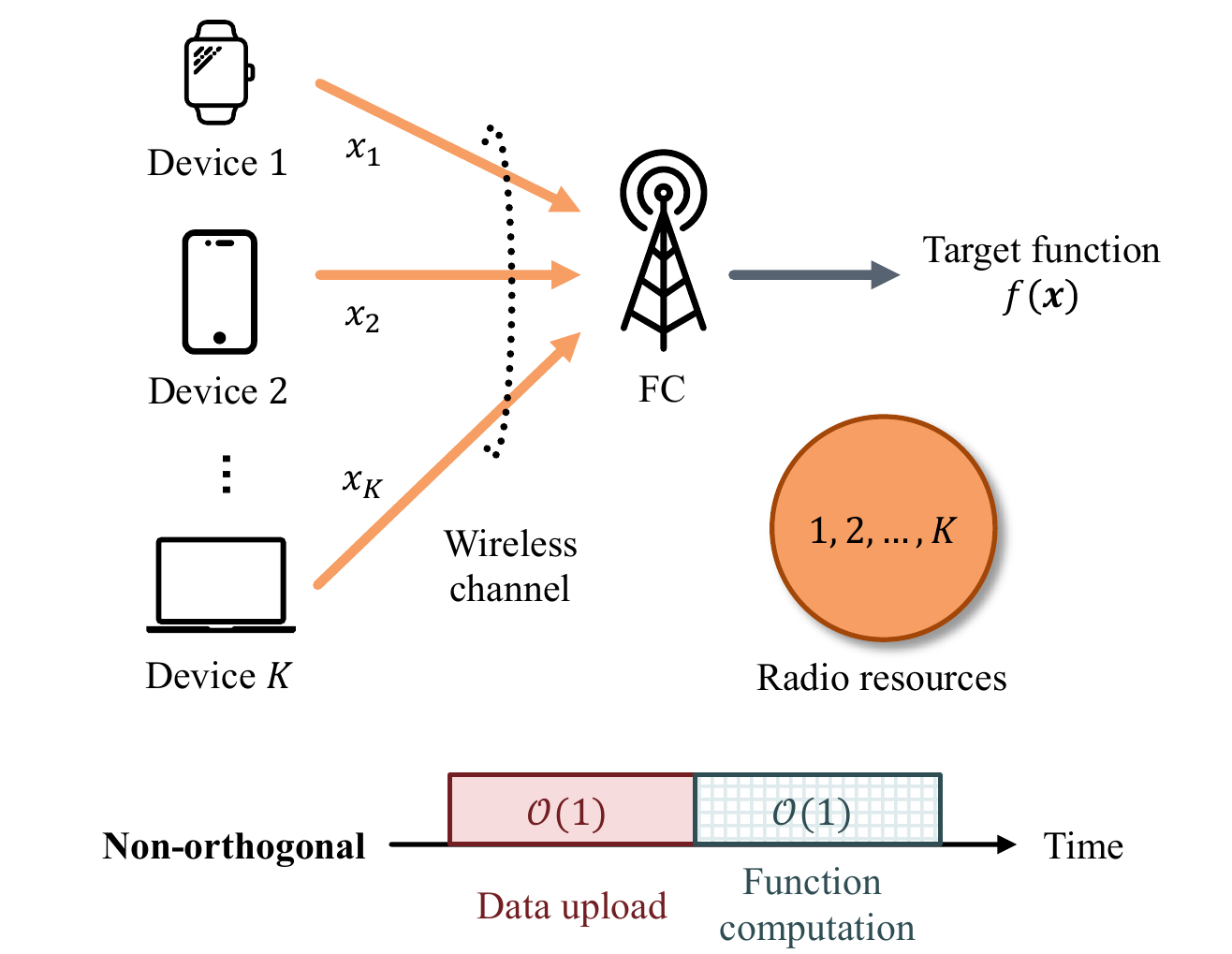}
	}
	\caption{Illustration of multiple-access strategies in wireless networks.} 
	\label{fig:aircomp_computation}
\end{figure}

\begin{table}[t]
	\centering
	\caption{``Compute-After-Communicate'' (\textsc{CaC}) v.s. ``Compute-When-Communicate'' (\textsc{CwC})}
	\label{tab:strategy}
	\scriptsize
	\renewcommand{\arraystretch}{1.5}
	\resizebox{\linewidth}{!}{
		\begin{tabular}{!{\vrule width1pt}>{\centering\arraybackslash}m{0.1\linewidth}!{\vrule width1pt}>{\centering\arraybackslash}m{0.13\linewidth}|m{0.4\linewidth}|>{\centering\arraybackslash}m{0.1\linewidth}!{\vrule width1pt}}
			\Xhline{1pt}
			\rowcolor[HTML]{e8e8e8}
			\bf{Strategy} & \bf{Resource allocation} & \multicolumn{1}{c|}{\bf{Procedure}} & \bf{Latency} \\ 
			\Xhline{1pt}
			\multirow{5}{*}{\textsc{CaC}} & Orthogonal & $\bullet$ Multiple access with $\mathcal{O}(K)$ resource blocks. \newline $\bullet$ Data recovery from orthogonal blocks with $\mathcal{O}(K)$ time. \newline $\bullet$ function computation based on $\mathcal{O}(K)$ recovered data. & \multirow{5}{*}{$\mathcal{O}(K)$} \\ 
			& \cellcolor[HTML]{f4f4f2} Non-orthogonal & \cellcolor[HTML]{f4f4f2} $\bullet$ Multiple access with $\mathcal{O}(1)$ resource blocks. \newline $\bullet$ Data recovery from superimposed signal via SIC with $\mathcal{O}(K)$ time. \newline $\bullet$ function computation based on $\mathcal{O}(K)$ recovered data. & \\ \hline
			\textsc{CwC} & Non-orthogonal & $\bullet$ Multiple access with $\mathcal{O}(1)$ resource blocks. \newline $\bullet$ function computation via processing $\mathcal{O}(1)$ superimposed signal. & $\mathcal{O}(1)$ \\
			\Xhline{1pt} 
		\end{tabular}
	}
\end{table}

\begin{figure*}[t]
	\centering
	\includegraphics[scale=0.55]{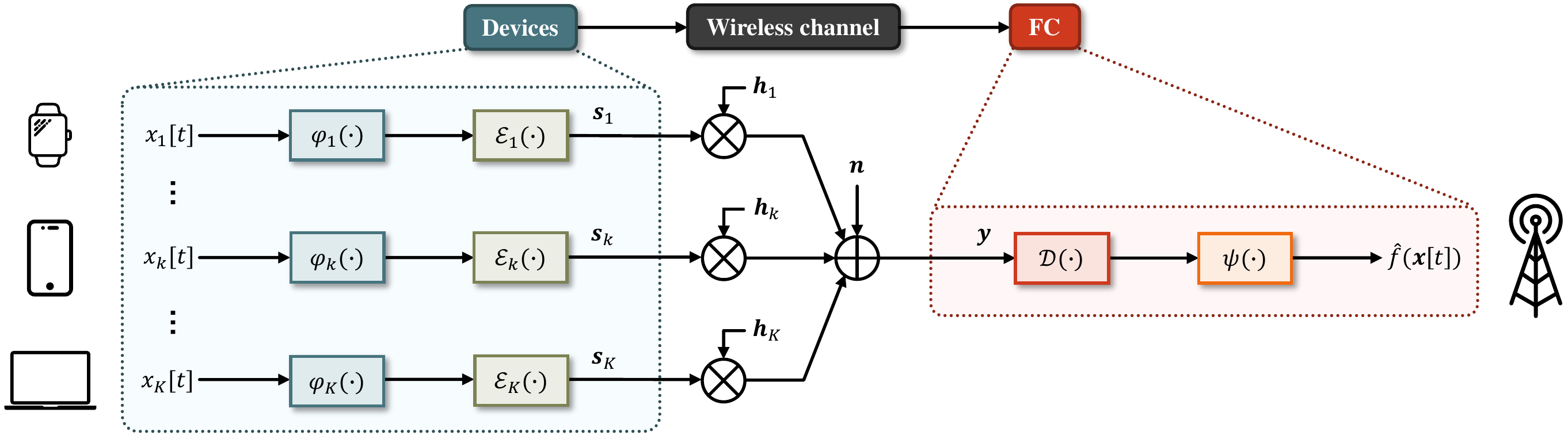}
	\vspace{-2mm}
	\caption{Workflow of AirComp. Specifically, each device $k \in \{1, 2, \dots, K\}$ first pre-processes its data $x_k[t] \in \mathbb{R}$ via function $\varphi_k(\cdot)$, $t \in \{1, 2, \dots, T\}$, and then converts the pre-processed data, $\{\varphi_k(x_k[t])\}$, into an $L$-length channel input signal, $\bm{s}_k = [s_{k, 1}, s_{k, 2}, \dots, s_{k, L}]^{\sf T}$, with encoder $\mathcal{E}_k(\cdot): \mathbb{R}^T \rightarrow \mathbb{C}^L$, where $L$ is a positive integer.
	Subsequently, all devices concurrently upload their signals $\{\bm{s}_k\}$ over MACs to execute a summation with each signal weighted by the channel coefficient.
	Herein, $\otimes$ realizes the element-wise multiplication, $\bm{h}_k \in \mathbb{C}^L$ denotes the channel coefficient vector of device $k$, and $\bm{n} \in \mathbb{C}^L$ is the additive noise.
	Eventually, the FC is capable of getting an estimate of the target functions, $\{\hat{f}(\bm{x}[t])\}$, by successively applying decoder $\mathcal{D}(\cdot):\mathbb{C}^L \rightarrow \mathbb{R}^T$ and post-processing $\psi(\cdot)$ to received superimposed signal $\bm{y} \in \mathbb{C}^L$, where $\bm{x}[t] = [x_1[t], x_2[t], \dots, x_K[t]]^{\sf T}$.}
	\label{fig:workflow}
\end{figure*}

In 1940s, the Shannon's landmark work \cite{shannon1948mathematical} presented a systematic analysis for point-to-point communications and proposed the Shannon-Hartley theorem for measuring the reliability of wireless data transmission.
Under this guidance, various OMA and NOMA communication schemes as mentioned in Section \ref{sec:intro} have been developed to guarantee an accurate reception of each message before the subsequent task processing.
In these schemes, the FC needs to successfully decode each of the individual signals from different devices before it can compute the desired function, which follows the principle of ``compute-after-communicate'', as shown in Fig. \ref{fig:comp_after_comm}.
Nevertheless, such paradigm isolates the communication process from the subsequent function computation, which inadvertently dissipates the performance gain that may be attained from the joint design.
AirComp addresses the above issue by adopting the ``compute-when-communicate'' strategy, as shown in Fig. \ref{fig:comp_when_comm}.
Instead of treating the concurrent signals as mutual troublemakers, AirComp harnesses the co-channel interference to compute the target function of delivered data.
Specifically, since the signals occupy the same radio channel for simultaneous data transmission, the number of required radio resources is independent of the number of connected devices, which is more spectral efficient than the OMA schemes that consumes the resource blocks in proportion to the number of devices.
Besides, as the signals can be naturally added over the air, the FC is able to obtain the desired function value by directly processing the received signal within one time slot in a symbol level, which reduces the delay of decoding individual signals one by one as in conventional OMA and NOMA schemes.
Moreover, by dividing target computation task $f(\cdot)$ into $K + 1$ distributed tasks $\{\psi(\cdot), \varphi_1(\cdot), \dots, \varphi_K(\cdot)\}$ according to Definition \ref{def:nomo_func}, the FC and each device merely need to perform lightweight signal processing themselves, which takes full advantage of the computing power in distributed devices while effectively reducing the computational complexity at the FC \cite{liu2020overtheair}.
It is also noteworthy that, if the target function needs to be decomposed into more than $K$ different basic nomographic functions for the calculation according to Theorem \ref{theorem:cont_func_nomo_func}, AirComp becomes spectral-inefficient as compared with conventional multiple-access schemes due to more than $K$ channel uses, which is the essential number of resources needed for OMA schemes, being required to accomplish the function computation \cite{zhu2021overtheair}.
The differences between ``compute-after-communication'' and ``compute-when-communicate'' these two strategies are highlighted in Table \ref{tab:strategy}.
For clarity, the workflow and communication design of AirComp is depicted in Fig. \ref{fig:workflow}.
With the aforementioned key differences between AirComp and conventional multiple-access schemes, we shall discuss three important aspects of AirComp that are critical for its communication design in the following.

\subsubsection{Taxonomy}

\begin{figure}[t]
	\centering
	\subfigure[Coded AirComp]{
		\centering
		\includegraphics[scale=0.55]{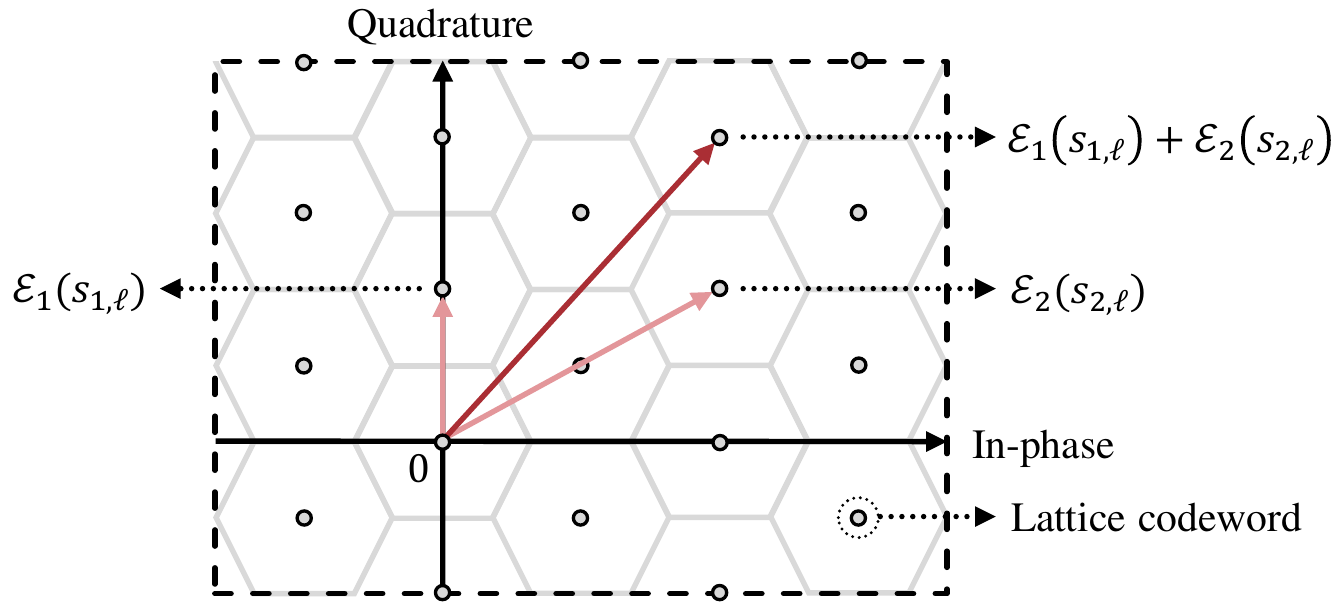}
		\label{fig:coded}
	}
	\subfigure[Uncoded AirComp]{
		\centering
		\includegraphics[scale=0.55]{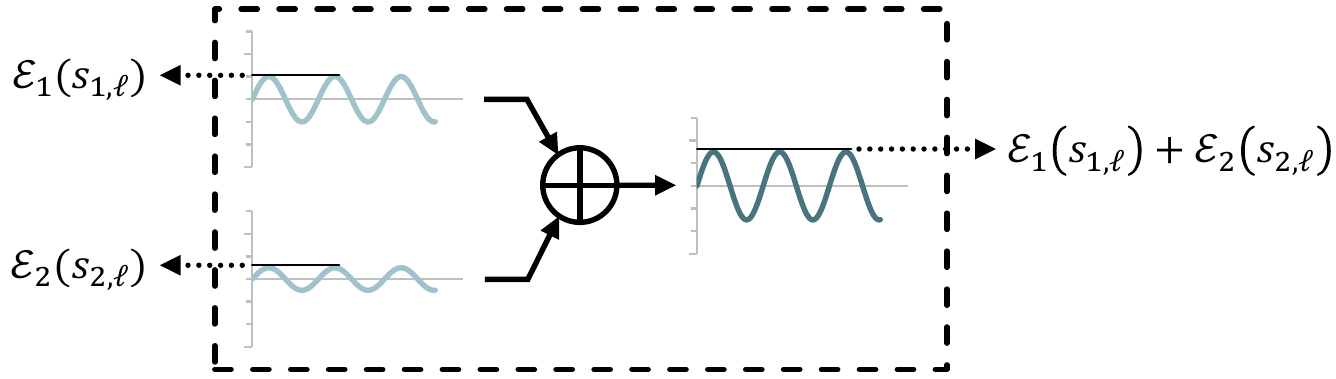}
		\label{fig:uncoded}
	}
	\caption{Examples of coded and uncoded AirComp with two devices.}
	\label{fig:comparison}
\end{figure}

To achieve the desired function computation over MACs, AirComp can be implemented via both coded and uncoded communications depending on the adopted modulation strategy, which are elaborated as follows.
\begin{itemize}
	\item \textbf{Coded AirComp:}
	Coding is an essential method for mitigating the signal distortion for conventional point-to-point communications, and is able to provide reliable transmissions by following the Shannon-Hartley theorem.
	Different from conventional coding strategies being used to combat interference, coded AirComp allows multiple devices to simultaneously transmit signals relying on the code with a linear structure, e.g., nested lattice code \cite{zamir2009lattices}, which enables the FC to reliably recover linear functions from the integer combinations of transmitted codewords \cite{nazer2011compute}, as shown in Fig. \ref{fig:coded}.
	Besides, an one-bit over-the-air aggregation method based on signSGD algorithm \cite{bernstein2018signsgd} has recently been proposed in \cite{zhu2021onebit}, which applies the one-bit quantization at the transmitters to simply represent the data via sign-taking operation and adopts a majority-voting based decoder at the receiver for estimating the sign of superimposed signals.
	By employing the above schemes, coded AirComp is able to combat the channel perturbation at the expense of reducing the computation accuracy due to the quantization error.
	Meanwhile, thanks to the intrinsic digital modulation, coded AirComp can be directly embedded into modern digital communication systems and integrated into the existing standards, e.g., 4G \textit{Long-Term Evolution} (LTE) and 5G \textit{New Radio} (NR).
	
	\item \textbf{Uncoded AirComp:}
	In contrast to the coded version that encodes data into discrete bits, uncoded AirComp directly modulates the information to the amplitude of the waveform of carrier signals, which can deliver high-precision data within a certain range of continuous values constrained by the transmit power, as shown in Fig. \ref{fig:uncoded}.
	By synchronizing the transmission of different devices, the signals transmitted over the same radio channel can be superimposed over the air with the weights being proportional to their corresponding channel coefficients.
	However, as the signal amplitude carries the transferred messages, uncoded AirComp is more fragile than the coded counterpart by directly exposing the carrier signals to the lossy and noisy wireless propagation, leading to unavoidable aggregation distortion at the FC.
	Therefore, it is critical for uncoded AirComp to implement an appropriate transceiver design to reduce the signal distortion and achieve the magnitude alignment at the receiver, thereby realizing the data aggregation with desired coefficients.
\end{itemize}

\subsubsection{Performance Metrics}

Since AirComp aims to achieve efficient function computation rather than the reliable transmission of each data, the computation rate is regarded as an important metric to evaluate the performance of AirComp systems \cite{nazer2007computation}. 
The definition of computation rate is given in the following.

\begin{definition}[Computation Rate \cite{goldenbaum2015nomographic, goldenbaum2015onachievable}] \label{def:comp_rate}
	Let an $(f, T, L)$ computation code for an MAC consist of: 1) A target function $f(\cdot): \mathbb{R}^K \rightarrow \mathbb{R}$; 2) An encoder at each of the $K$ transmitters to map $T$ generated data $\{x_k[t]\}$ to $L$ channel input symbols $\{s_{k, \ell}\}$ such that $\left(x_k[1], x_k[2], \dots, x_k[T]\right) \mapsto \left(s_{k, 1}, s_{k, 2}, \dots, s_{k, L}\right), \, \forall \, k \in \{1, 2, \dots, K\}$; and 3) A decoder at the receiver to obtain $T$ estimates of the target function based on $L$ channel output symbols $\{y_\ell\}$ such that $\left(y_1, y_2, \dots, y_L\right) \mapsto (\hat{f}(\bm{x}[1]), \hat{f}(\bm{x}[2]), \dots, \hat{f}(\bm{x}[T]))$ with $\bm{x}[t] = \left[x_1[t], x_2[t], \dots, x_K[t]\right]^{\sf T}$.
	Then, given an arbitrary but a fixed computation error requirement, $\epsilon > 0$, computation rate $R_{\rm C}$, specifying the number of computed function values per channel use, is achievable if, for every rate $R = \frac{T}{L} \le R_{\rm C}$ and $\delta \in (0, 1)$, there exists a sequence of $(f, T, L)$ computation codes such that, for sufficiently large $L$, $\mathbb{P}\left[\bigcup_{t = 1}^T \{|\hat{f}(\bm{x}[t]) - f(\bm{x}[t])| \ge \epsilon\}\right] \le \delta$ for coded AirComp or $\mathbb{E}\left[|\hat{f}(\bm{x}[t]) - f(\bm{x}[t])|^2\right] \le \epsilon$, $\forall \, t$, for uncoded AirComp.
\end{definition}

Compared with the maximum reliable transmission rate quantifying the bits successfully transmitted per second, the computation rate specifies the maximum number of functions that can be computed with a single channel usage under a desired computation accuracy, regardless of the reliability of individual signal transmissions.
Note that the computation code given in Definition \ref{def:comp_rate} represents a signal mapping relationship, which can be either a digital coding that encodes the signal into multiple binary bits or an analog processing that transmits a power-scaling signal over multiple channel blocks.
In addition, the computation accuracy is another critical performance metric and is generally measured by the \textit{mean-squared error} (MSE), i.e., $\mathbb{E}\left[|\hat{f}(\bm{x}) - f(\bm{x})|^2\right]$, which quantifies the distortion between the estimated and the ground-truth values of the target function.

\subsubsection{Synchronization}

Precise time and frequency synchronization among multiple devices is critical for ensuring that the signals transmitted from multiple devices can simultaneously arrive at the receiver, thereby realizing the accurate function computation over the air.
Fortunately, various well-studied synchronization techniques can be adopted to achieve synchronization for AirComp.
For instance, the timing advance technique is commonly used in 4G LTE and 5G NR to achieve time synchronization by controlling the uplink transmission timing of each device \cite{mahmood2019time}.
Specifically, each device first evaluates its propagation delay through the timestamps of the reference time broadcast by the FC, and then compensates for the corresponding delay by advancing or retarding its uplink transmission.
This ensures that the signals transmitted from all devices can simultaneously arrive at the FC.
In addition, frequency synchronization is also required in AirComp to eliminate \textit{carrier frequency offset} (CFO) in concurrently transmitted signals, which is critical for combining the signals at the FC in a desired manner.
To this end, a primitive method, named AirShare, is proposed in \cite{abari2015airshare} to realize distributed coherent transmissions by sharing a reference-clock with multiple devices to eliminate the CFO by synchronizing their generating clocks of crystal oscillators.
Consequently, the above synchronization techniques enable AirComp to be implemented in practical wireless networks.


\section{AirComp over Different Network Architectures}\label{sec:network_architecture}

Diverse applications impose different requirements on the function computation, such as low-latency, multi-target, and multi-tiered computation, leading to the exploration of different network architectures to tackle these issues.
In this section, we review the existing studies on AirComp over various wireless network architectures, including single-cell, multi-cell, hierarchical, decentralized, \textit{reconfigurable intelligent surface} (RIS)-aided, and \textit{unmanned aerial vehicle} (UAV)-aided networks, and discuss the major issues therein.
The outline of this section can be previewed in Table \ref{tab:net}.

\begin{table*}[t]
	\centering
	\caption{Outline of AirComp over Different Network Architectures}
	\label{tab:net}
	\scriptsize
	\renewcommand{\arraystretch}{1.5}
	\resizebox{\linewidth}{!}{
	\begin{tabular}{!{\vrule width1pt}c!{\vrule width1pt}>{\centering\arraybackslash}m{0.137\linewidth}|>{\centering\arraybackslash}m{0.07\linewidth}|>{\centering\arraybackslash}m{0.13\linewidth}|m{0.46\linewidth}!{\vrule width1pt}}
		\Xhline{1pt}
		\rowcolor[HTML]{e8e8e8}
		\multicolumn{1}{!{\vrule width1pt}c!{\vrule width1pt}}{\bf{Architecture}} & \multicolumn{1}{c|}{\bf{Feature}} & \multicolumn{1}{c|}{\bf{Reference}} & \multicolumn{1}{c|}{\bf{Performance criterion}} & \multicolumn{1}{c|}{\bf{Tackled issue}} \\ \Xhline{1pt}
		\multirow{6}{*}{\shortstack{Single-cell \\ network}} & \multirow{6}{*}{\begin{tabular}
				{@{}m{\linewidth}@{}}Single function computation task by harnessing intra-cell interference.
		\end{tabular}} \raggedright & \cite{nazer2007computation, nazer2011compute, goldenbaum2015nomographic, jeon2016opportunistic, chen2019communicating, chen2021toward} & Computation rate & Computation rate analyses for coded AirComp over different types of MACs. \\
		& & \cellcolor[HTML]{f4f4f2} \cite{liu2020overtheair, cao2020optimized, zang2020overtheair, chen2018uniform, chen2018overtheair, zhu2019mimo, wen2019reduced} & \cellcolor[HTML]{f4f4f2} Computation error & Transceiver design for narrowband transmission. \cellcolor[HTML]{f4f4f2} \\
		& & \cite{wu2019computation, wu2020nomaenhanced, qin2021overtheair, kulhandjian2021noma} & Computation rate & Computation rate analyses and transceiver design for wideband transmission. \\
		& & \cellcolor[HTML]{f4f4f2} \cite{liu2021spatialtemporal, nakai2022precoder, frey2021corrchannel} & \cellcolor[HTML]{f4f4f2} Computation error & Transceiver design for AirComp with spatial and temporal correlated signals.  \cellcolor[HTML]{f4f4f2} \\
		& & \cite{frey2021corrchannel} & Computation error & Theoretical performance analyses for AirComp with correlated channels. \\ \hline
		\multirow{4}{*}{\shortstack{Multi-cell \\ network}} & \multirow{4}{*}{\begin{tabular}
				{@{}m{\linewidth}@{}}Multiple function computation tasks in different cells by managing both intra- and inter-cell interference.
		\end{tabular}} \raggedright & \cellcolor[HTML]{f4f4f2} \cite{goldenbaum2015nomographic} & \cellcolor[HTML]{f4f4f2} Computation rate & \cellcolor[HTML]{f4f4f2} Computation rate analyses for multi-cluster AirComp with time-divisional computations. \\
		& &\cite{lan2020simultaneous} & Degrees-of-freedom & \textit{Signal-and-interference alignment} (SIA) precoding design for two-cell AirComp. \\
		& & \cellcolor[HTML]{f4f4f2} \cite{cao2021cooperative, zhang2022interference, li2023multicell} & \cellcolor[HTML]{f4f4f2} Computation error & \cellcolor[HTML]{f4f4f2} Interference management and transceiver design for multi-cell AirComp with centralized/decentralized implementation. \\ \hline
		\multirow{9}{*}{\shortstack{Hierarchical \\ network}} & \multirow{9}{*}{\begin{tabular}
				{@{}m{\linewidth}@{}}Single function computation task over a large-scale network by performing multi-level data aggregations.
		\end{tabular}} \raggedright & \cite{wang2022amplifyforward, jiang2021joint, li2022joint} & Computation error & Transceiver design for \textit{amplify-and-forward} (AF)-relay aided hierarchical AirComp. \\
		& & \cellcolor[HTML]{f4f4f2} \cite{jiang2020achieving} & \cellcolor[HTML]{f4f4f2} Computation error & Relay selection design for AF-relay assisted hierarchical AirComp. \cellcolor[HTML]{f4f4f2} \\
		& & \cite{tang2021reliable, tang2022node} & Computation error and power consumption & Device scheduling design for achieving energy-efficient AF-relay assisted hierarchical AirComp. \\
		& & \cellcolor[HTML]{f4f4f2} \cite{yu2020optimizing, xing2021overtheair} & \cellcolor[HTML]{f4f4f2} Computation error & Transceiver design and fronthaul link quantization bits allocation for \textit{cloud radio access network} (Cloud-RAN) assisted hierarchical AirComp. \cellcolor[HTML]{f4f4f2} \\
		& & \cite{wu2021computation} & Computation rate & Computation rate analyses for multi-hop AirComp with group and subgroup constructions. \\
		& & \cellcolor[HTML]{f4f4f2} \cite{wang2022multilevel} & \cellcolor[HTML]{f4f4f2} Computation error & Transceiver design for multi-hop AirComp constructed by decentralized devices via \textit{device-to-device} (D2D) communications. \cellcolor[HTML]{f4f4f2} \\ \hline
		\multirow{6}{*}{\shortstack{Decentralized \\ network}} & \multirow{6}{*}{\begin{tabular}
				{@{}m{\linewidth}@{}}Data consensus over distributed devices without centralized coordinator.
		\end{tabular}} \raggedright & \cite{nazer2011local} & Computation rate, consensus time, and energy consumption & Gossip algorithm acceleration with decentralized AirComp. \\
		& & \cellcolor[HTML]{f4f4f2} \cite{zheng2012fast, goldenbaum2012nomographic} & \cellcolor[HTML]{f4f4f2} Consensus rate & \cellcolor[HTML]{f4f4f2} Multi-cluster data consensus through the common nodes located in overlapping areas of different clusters. \\
		& & \cite{molinari2018exploiting, molinari2020exploiting, molinari2021maxconsensus} & Consensus rate & Data consensus without prior knowledge of channel conditions. \\
		& & \cellcolor[HTML]{f4f4f2} \cite{ozfatura2020decentralized, shi2021overtheair, xing2021federated} & \cellcolor[HTML]{f4f4f2} Computation error & \cellcolor[HTML]{f4f4f2} Transceiver design for decentralized AirComp assisted FL. \\
		& & \cite{lin2023distributed} & Computation error & Transceiver design for distributed optimization with \textit{full-duplex} (FD) devices. \\ \hline
		\multirow{4}{*}{\shortstack{RIS-aided \\ network}} & \multirow{4}{*}{\begin{tabular}
				{@{}m{\linewidth}@{}}Reliable data aggregation with reconfigured wave propagation environment by establishing reflective paths with RIS. 
		\end{tabular}} \raggedright & \cellcolor[HTML]{f4f4f2} \cite{jiang2019overtheair, fang2021overtheair} & \cellcolor[HTML]{f4f4f2} Computation error & \cellcolor[HTML]{f4f4f2} Preliminary studies on transceiver and phase shifts design for RIS-aided AirComp. \\
		& & \cite{li2021doubleris, zhai2022joint} & Computation error & Transceiver and phase shifts design for double-RIS aided AirComp. \\
		& & \cellcolor[HTML]{f4f4f2} \cite{zhai2022beamforming} & \cellcolor[HTML]{f4f4f2} Computation error & \cellcolor[HTML]{f4f4f2} Two time-scale transceiver and phase shifts design for RIS-aided AirComp. \\
		& & \cite{li2022reconfigurable} & Computation error & Transceiver and phase shifts design for RIS-aided multi-cell AirComp. \\ \hline
		\multirow{8}{*}{\shortstack{UAV-aided \\ network}} & \multirow{8}{*}{\begin{tabular}
				{@{}m{\linewidth}@{}}Flexible data aggregation with extensive coverage area by resorting to the mobility of UAVs.
		\end{tabular}} \raggedright &\cellcolor[HTML]{f4f4f2}  \cite{farajzadeh2020mobility} & \cellcolor[HTML]{f4f4f2} Computation error & \cellcolor[HTML]{f4f4f2} Power allocation for UAV-aided AirComp in backscatter sensor networks. \\
		& &  \cite{joung2021overtheair} & Computation error & Space-time line code design for UAV-aided AirComp. \\
		& & \cellcolor[HTML]{f4f4f2} \cite{fu2021uav} & \cellcolor[HTML]{f4f4f2}  Computation error & \cellcolor[HTML]{f4f4f2}  Transceiver and trajectory design for UAV-aided AirComp with mobile devices. \\
		& &  \cite{zeng2022optimized} & Computation error & Transceiver and trajectory design for UAV-aided AirComp with large-scale distributed devices. \\
		& & \cellcolor[HTML]{f4f4f2}  \cite{fu2023uav} & \cellcolor[HTML]{f4f4f2} Computation error & \cellcolor[HTML]{f4f4f2} Transceiver and trajectory design for multiple UAVs aided multi-cluster AirComp. \\
		& & \cite{jung2022performance} & Computation error & Theoretical performance analyses for UAV-aided AirComp with channel and synchronization errors. \\
		\Xhline{1pt} 
	\end{tabular}
	}
	\vspace{-3mm}
\end{table*}

\subsection{Single-Cell Network} \label{subsec:single_cell}

\begin{figure}[t]
	\centering
	\includegraphics[scale=0.55]{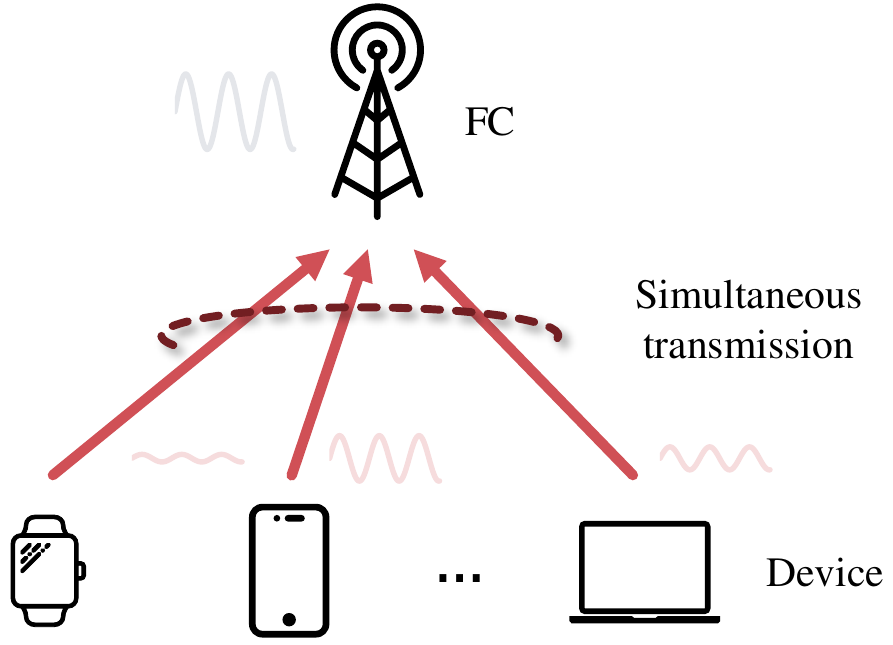}
	\caption{AirComp over a single-cell network.}
	\label{fig:singlecell}
\end{figure}

For AirComp over canonical single-cell networks, as shown in Fig. \ref{fig:singlecell}, the FC aims to aggregate the signals concurrently transmitted from multiple devices, where the intra-cell interference is harnessed to compute the desired function through the waveform superposition over MACs.
Accordingly, how to effectively exploit the intra-cell interference to achieve efficient and accurate function computation becomes the main focus of the research on single-cell AirComp, while the single-cell algorithmic explorations can also provide insights for extending the associated designs to other network architectures.

To this end, digital linear-structured codes and analog transceiver design are developed to achieve desired computation with coded and uncoded AirComp, respectively.
For instance, linear-structured coding strategy are considered in \cite{nazer2007computation, nazer2011compute, goldenbaum2015nomographic, jeon2016opportunistic, chen2019communicating, chen2021toward} to enable the quantized data bits to be superimposable via nested lattice codes, which is capable of combating the channel noise by correcting the disturbed overlapped signals back to the nearest lattice point \cite{goldenbaum2015nomographic}.
Based on this, achievable computation rates of coded AirComp are analyzed under different channel models, such as discrete linear \cite{nazer2007computation}, Gaussian \cite{nazer2007computation, nazer2011compute}, constant \cite{goldenbaum2015nomographic}, and fading \cite{jeon2016opportunistic, chen2019communicating} MACs.
Besides, in order to superimpose analog signals over the air with desired coefficients, transceiver design for various scenarios is studied to align the magnitudes and phase shifts of received concurrent signals at the FC, such as optimal power control for \textit{single-input single-output} (SISO) systems \cite{liu2020overtheair, cao2020optimized, zang2020overtheair} and uniform-forcing beamforming for \textit{single-input/multiple-input multiple-output} (SIMO/MIMO) systems \cite{chen2018uniform, chen2018overtheair, zhu2019mimo, wen2019reduced}.
In addition to narrowband transmissions, wideband AirComp is investigated in \cite{wu2019computation, wu2020nomaenhanced, qin2021overtheair, kulhandjian2021noma} to fully utilize the available bandwidth resources for data aggregation while overcoming \textit{inter-symbol interference} (ISI) through the integration of narrowband AirComp and \textit{orthogonal frequency division multiplexing} (OFDM) technique.
With broadband AirComp, parallel function computation can be achieved by overlapping the signals transmitted over different subchannels.
Moreover, the authors in \cite{liu2021spatialtemporal} and \cite{nakai2022precoder} incorporate the statistics of data correlation among different devices and time slots with the transceiver design for AirComp, where the spatial and temporal information of transmit signals is exploited for further computation error reduction in addition to the \textit{channel state information} (CSI) across different devices.
Furthermore, the authors in \cite{frey2021corrchannel} derive non-asymptotical computation error bounds for AirComp, which are applicable to the scenarios with fast-fading/sub-Gaussian/correlated channels and independent/correlated signals.

\subsection{Multi-Cell Network}

\begin{figure}[t]
	\centering
	\includegraphics[scale=0.55]{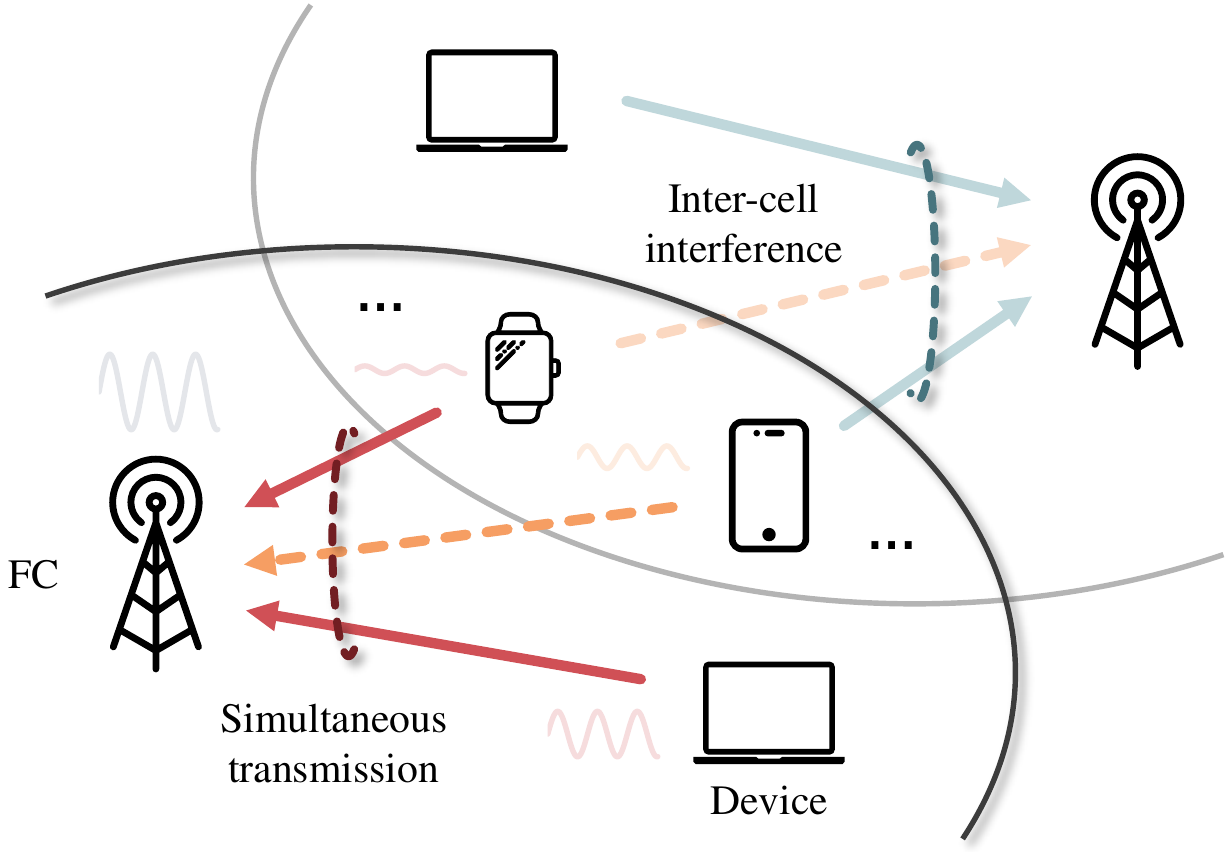}
	\caption{AirComp over a multi-cell network.}
	\label{fig:multicell}
\end{figure}

Benefiting from high spectral efficiency and low communication latency, AirComp is promising to enable a large-scale wireless data aggregation, which necessitates the multi-cell network to capacitate extensive connections while supporting the computation of different target functions characterized by different services.
Similar to the co-existence of multiple services in single-cell networks, the multi-cell setting requires interference management for dealing with multiple co-existed AirComp tasks in different cells, as shown in Fig. \ref{fig:multicell}, such that utilizing intra-cell interference within each cell to support function computation and alleviating inter-cell interference among different cells to reduce the computation distortion.

Interference management for conventional multi-cell wireless networks has been studied for many years \cite{cadambe2008interference, gesbert2010multicell, zhang2010cooperative, dahrouj2010coordinated}.
An intuitive strategy is to avoid interference through orthogonal resource allocation, which has been exploited in \cite{goldenbaum2015nomographic} for the multi-cluster function computation by activating each cluster in a time-division manner.
Besides, among the emerging interference management techniques, \textit{interference alignment} (IA) is a promising one to realize high transmission rates in interference channels by dividing the MIMO channel space into two subspaces for distinguishing the interference and desired signals \cite{cadambe2008interference}.
Motivated by this, the authors in \cite{lan2020simultaneous} present a simultaneous SIA scheme for a two-cell AirComp system.
Unlike the conventional IA scheme demanding an array size scaling with the the numbers of signal streams and devices \cite{cadambe2008interference}, the developed SIA scheme in \cite{lan2020simultaneous} performs a symmetric subspace division regardless of the number of devices in multi-cell AirComp systems, which requires only twice as many antenna arrays as spatially multiplexing streams of MIMO AirComp in each cell.
Besides, the authors in \cite{cao2021cooperative} present a centralized cooperative power control scheme to minimize the weighted-sum MSE of multi-cell AirComp and a distributed optimization method to further reduce the communication overhead, in which the multi-cell power control problem reduces to separated single-cell optimizations by introducing interference temperature constraints that limits the inter-cell interference from one cell to another one.
Moreover, the authors in \cite{zhang2022interference} study the interference management for multi-cell AirComp with multi-antenna FCs, which reveals that, when the number of receive antennas tends to be infinity, the asymptotically optimal receive beamforming of each FC in each cell can be approximated as a linear combination of the channel coefficients between the FC itself and its served devices.
In addition, the authors in \cite{li2023multicell} propose a unified optimization framework for multi-cell SIMO AirComp from the perspective of the fairness of data aggregation.
Under the SIMO AirComp system, the asymptotic analyses in \cite{zhang2022interference} and \cite{li2023multicell} both indicate that the distortion of AirComp caused by inter-cell interference can be eliminated when there is a large enough antenna array size at the FC, in which case the single-cell design comes back into play.

\subsection{Hierarchical Network}

As stated in the above subsection, multi-cell network is a remedy to tackle multiple co-existed services in large-scale networks.
If a common task needs to be accomplished in a large-scale network, the hierarchical network with a multi-hop topology is preferable to be leveraged to address the challenges posed by single-hop networks with direct communications, such as the requirement for precise synchronization of all devices and the compensation for severe propagation loss caused by long communication distances.

\begin{figure}[t]
	\centering
	\includegraphics[scale=0.55]{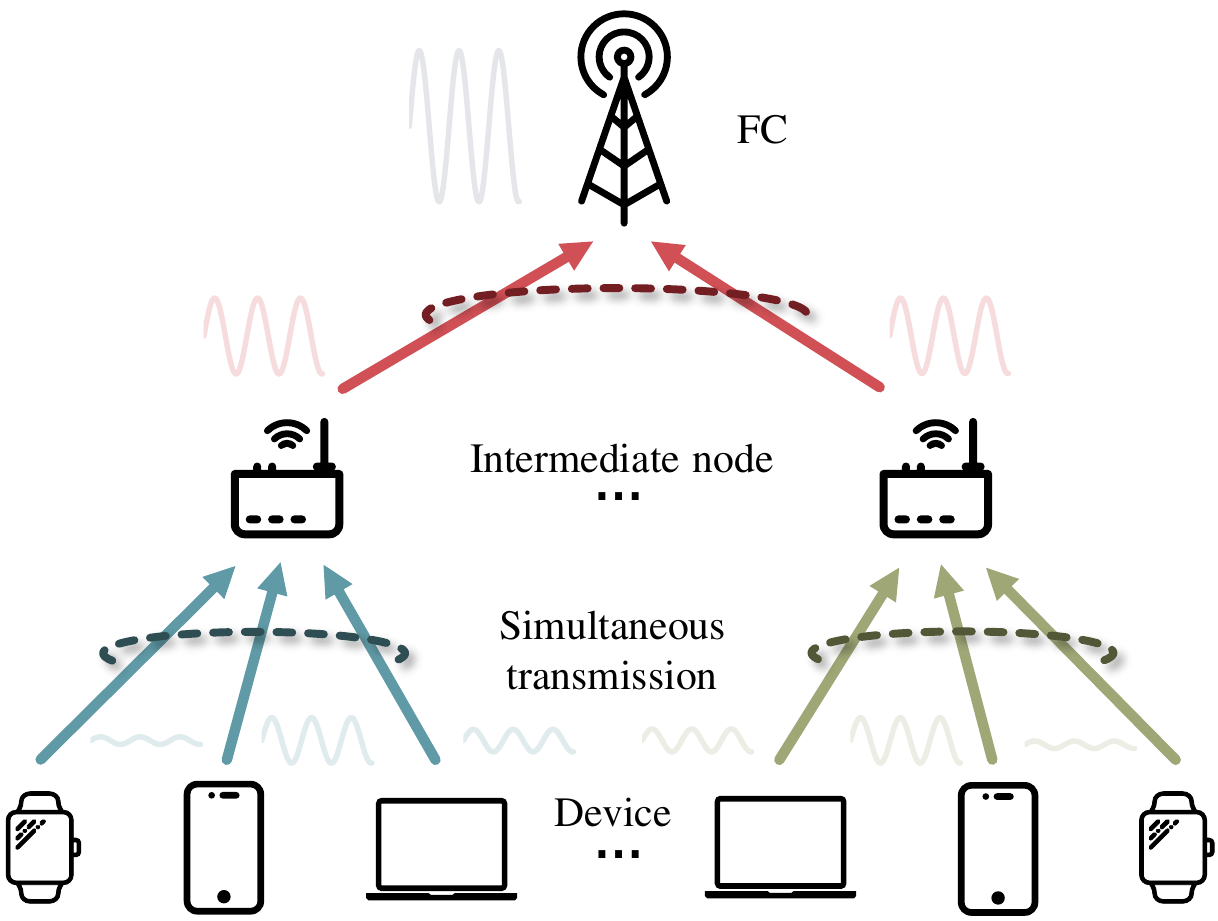}
	\caption{AirComp over a hierarchical network.}
	\label{fig:hierarchical}
\end{figure}

Recently, several works have exploited the hierarchical architecture to promote the performance of AirComp in large-scale networks, as shown in Fig. \ref{fig:hierarchical}.
Specifically, the authors in \cite{wang2022amplifyforward} develop a two-phase AF relaying protocol for hierarchical AirComp systems, where the first-phase data aggregation from the devices to the relays and second-phase data aggregation from the relays to the AP are both realized via AirComp.
To minimize the computation error, both the centralized and decentralized design with global and partial CSI, respectively, are considered in \cite{wang2022amplifyforward}.
The derived solutions to the centralized design necessitate the adjustment of the phase of transmit coefficient of each device to be an opposite phase of the device-relay-FC channel and the transmit power of each device/relay to be a regularized composite-channel-inversion structure, such that ensuring a phase alignment while mitigating the magnitude misalignment error at the FC.
As for the decentralized design, the transmit coefficients of relays are modified solely relying on the local CSI by regarding the inter-relay signals as detrimental interference or noise, which effectively mitigates the signaling overhead for coordinating multiple relays with global CSI.
Besides, the authors in \cite{jiang2021joint} and \cite{li2022joint} study the beamforming design for AirComp with multi-antenna relays by considering a two-phase communication with both direct and relay links, which demonstrates that additional performance gains can be obtained from the direct links.
Instead of exploiting all relays deployed in the network, the authors in \cite{jiang2020achieving} propose a relay selection scheme based on the source-relay and relay-FC channel conditions, where only one relay is scheduled for forwarding the data.
By adopting the channel inversion design, the MSE outage probability of the relay selection scheme derived by \cite{jiang2020achieving} achieves the same diversity order of the SIMO AirComp with antenna selection, but suffering a \textit{signal-to-noise ratio} (SNR) loss as compared with the SIMO setting.
To realize energy-efficient hierarchical AirComp, the authors in \cite{tang2021reliable} propose to only allow a subset of devices with small differences in channel gains between device-FC and device-relay to employ the associated relay for the data delivery, while the others conduct direct communications with the FC. The presented relay policy is able to yield lower power consumption and smaller MSE as compared with the AirComp without the assistance of relay, but at the cost of one more communication slot.
Furthermore, device scheduling methods are also developed in \cite{tang2022node} under the power constraint of relay, which significantly reduces the computation MSE with little extra power consumption as compared with the approach proposed in \cite{tang2021reliable}.

Moreover, the authors in \cite{yu2020optimizing} and \cite{xing2021overtheair} resort to the Cloud-RAN architecture for hierarchical AirComp, where the devices first transmit signals to distributed \textit{remote radio heads} (RRHs) via AirComp and the RRHs then forward the aggregated data to the baseband unit through the capacity-limited fronthaul links.
In addition, the authors in \cite{wu2021computation} propose a multi-layer AirComp system for composite function computation by dividing nodes into groups and further into subgroups in each layer.
With the nested lattice coding, theoretical analysis provided in \cite{wu2021computation} indicates that the computation rate of multi-hop AirComp is bottlenecked by the worst rate of computing subgroup functions, suggesting the low-rate subgroup to be allocated more channel uses than the high-rate subgroups for increasing the minimum number of computed subfunctions.
Besides, the authors in \cite{wang2022multilevel} construct a multi-hop AirComp based on D2D communications, where the devices can be designated as the intermediate nodes of surrounding areas during multi-hop transmissions.
To reduce the aggregation error accumulated over different levels, the multi-hop AirComp in \cite{wang2022multilevel} is modeled as a \textit{minimum spanning tree} (MST) that has a smallest level number of the connected graph topology, which yields a smaller MSE than the non-MST multi-hop topology.

\subsection{Decentralized Network}

\begin{figure}[t]
	\centering
	\includegraphics[scale=0.55]{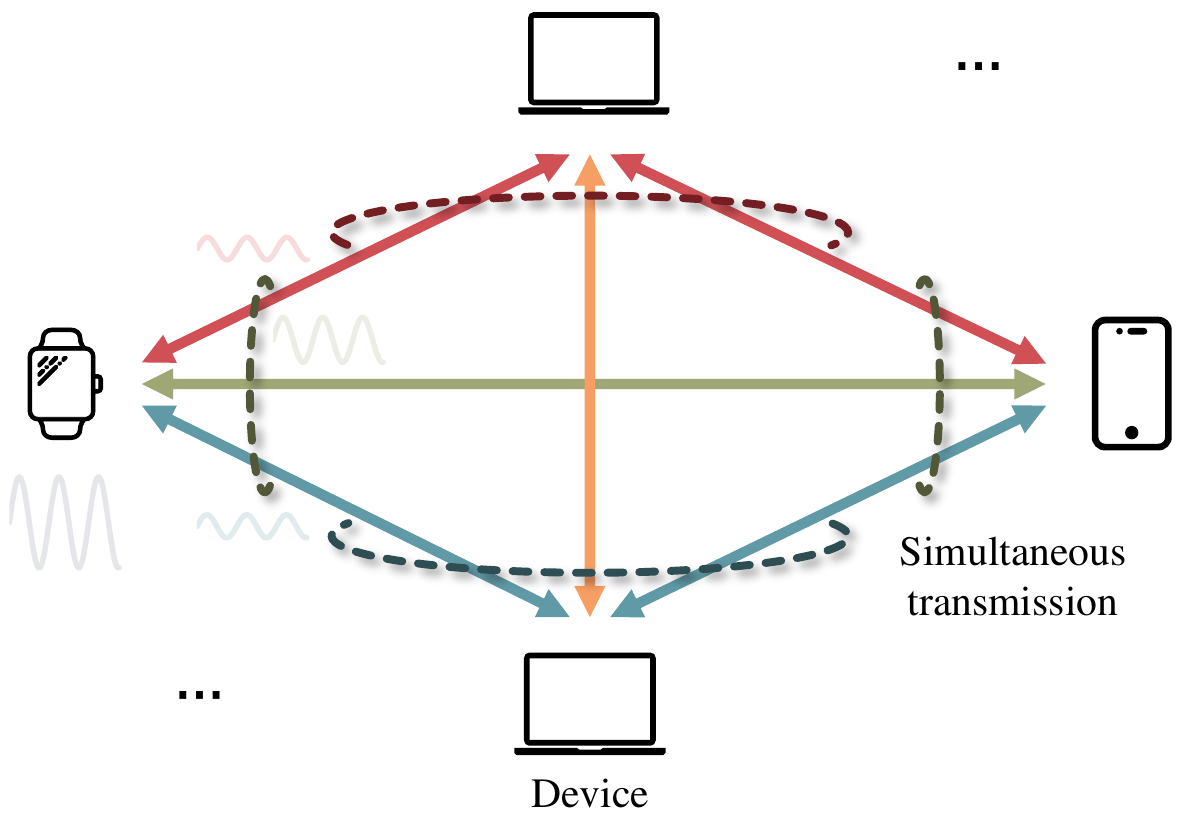}
	\caption{AirComp over a decentralized network.}
	\label{fig:decentralized}
\end{figure}

In some application scenarios, e.g., cooperative autonomous vehicles and swarm robotics, the behaviors of devices are characterized by time-sensitive functions, which require the corresponding function computation to be finished within ultra-low latency \cite{savazzi2021opportunities}.
This makes it much intractable to rely on a centralized server to coordinate massive dispersed devices in such scenarios. 
Besides, the performance of a centralized system is generally hampered by the straggler devices with poor communication resources, which prolongs the communication delay and reduces the transmission reliability.
Therefore, decentralized system with D2D communications is preferred to tackle these issues, which enables devices to communicate with their neighbors through strong and reliable wireless links \cite{doppler2009devicetodevice, fodor2012design, asadi2014survey}.

Motivated by this, decentralized AirComp is studied for fast data consensus to get rid of the dependence on a centralized FC, as shown in Fig. \ref{fig:decentralized}.
For instance, the authors in \cite{nazer2011local} leverage AirComp to accelerate the gossip algorithm for average-based data consensus.
By allowing each device to compute an average of multiple neighboring observations at once, the AirComp-enabled gossip algorithm reduces the number of consensus rounds required for convergence as compared to the conventional gossip with pairwise averaging \cite{nazer2011local}.
Meanwhile, the energy efficiency is improved by inducing a polynomially lower energy consumption than the standard gossip when the path loss coefficient is less than four \cite{nazer2011local}.
Besides, a network-wide consensus over a multi-cluster wireless network is considered in \cite{zheng2012fast} and \cite{goldenbaum2012nomographic}, where AirComp is utilized to enable one-shot data aggregation within each cluster.
Instead of assuming ideal/known channel coefficients, the authors in \cite{molinari2018exploiting} develop an AirComp-assisted consensus protocol in a blind manner without accessing the channel conditions.
By letting each device broadcast two orthogonal signals, i.e., one is the data to be consensus and the other is value one, to neighbor devices, the consensus update matrix at each device can be formulated as a row-stochastic matrix based on the two types of aggregated data, allowing for eventual data consensus without prior CSI knowledge \cite{molinari2018exploiting, molinari2020exploiting, molinari2021maxconsensus}.
In addition, the authors in \cite{ozfatura2020decentralized, shi2021overtheair, xing2021federated} present power control schemes for decentralized AirComp, where the uniform-forcing transmit scalar design is adopted toward an unbiased aggregation at each receiver.
A noise-aware aggregation scheme is further proposed in \cite{ozfatura2020decentralized} by additionally introducing a constraint for the receive scaling factor, based on which the aggregation matrix of each device is adjusted to further suppress the receiver noise.
Moreover, the authors in \cite{lin2023distributed} develop a decentralized AirComp framework with FD devices to achieve efficient distributed optimization, which significantly reduces the communication latency as compared with \textit{half-duplex} (HD) scenarios by paralleling the aggregation at different devices.

\subsection{RIS-Aided Network}

\begin{figure}[t]
	\centering
	\includegraphics[scale=0.55]{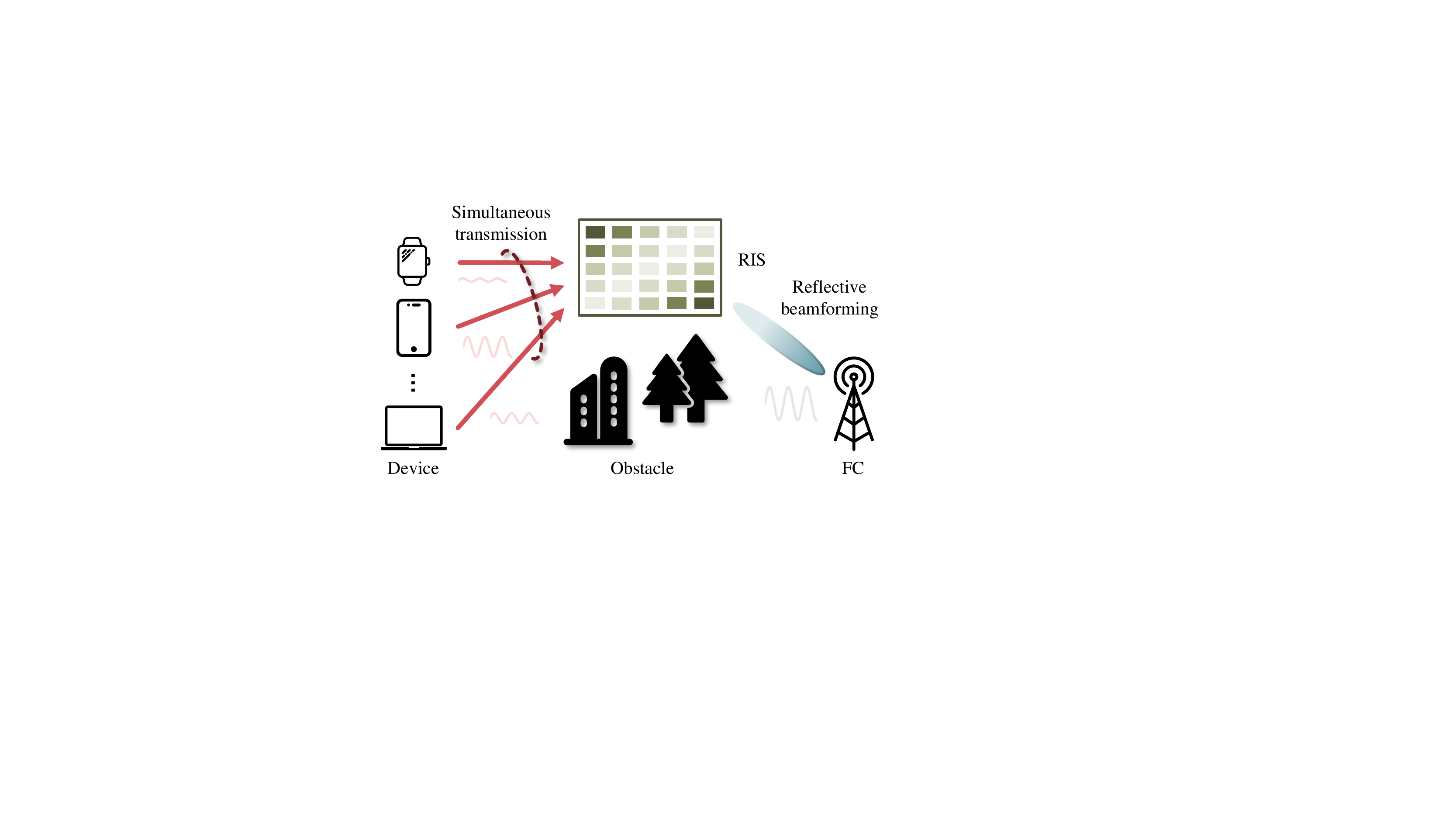}
	\caption{AirComp over an RIS-aided network.}
	\label{fig:ris}
\end{figure}

The uncontrollable propagation environment is deemed as an unintentional adversary to wireless communications, and only the transceiver design at the end nodes can be optimized to mitigate the adverse impact caused by the environment, e.g., channel deep fading \cite{tse2005fundamentals}.
Besides, the radio propagation is usually blocked by the obstacles like buildings and hills, which leads to significant energy decaying on signals, especially in high frequencies, e.g., millimeter and terahertz waves \cite{mezzavilla2018endtoend, ghafoor2020mac}.
To overcome these challenges, RIS emerges as a cost-effective technique for enhancing the reliability and efficiency of wireless communications via reconfiguring the wave propagation environment \cite{wu2021tutorial, yuan2021reconfigurable}.
An RIS typically consists of several reflecting elements, which are capable of adjusting the phases of incident signals through an embedded controller, thereby generating reflective beamforming for boosting the received signal power.

Inspired by this new communication paradigm, RIS-aided AirComp has been studied in several works to further ameliorate the performance of AirComp, as shown in Fig. \ref{fig:ris}.
Specifically, the authors in \cite{jiang2019overtheair} and \cite{fang2021overtheair} make preliminary studies for RIS-assisted AirComp, which show a prominent improvement on computation accuracy by leveraging the RIS to promote the worst channel gain among all devices.
Then, double-RIS designs are developed in \cite{li2021doubleris} and \cite{zhai2022joint} to further reduce computation error of AirComp through the coordination of single- and double-reflection links.
Due to the concern of excessive overhead for controlling the phase shifts at the RIS, the authors in \cite{zhai2022beamforming} propose a two-stage beamforming design to minimize the time-average MSE of AirComp.
In particular, the transmit power and receive beamforming are optimized at each time slot based on real-time low-dimensional cascaded CSI, while the phase shifts at the RIS is updated once every few slots based on the channel statistics, thereby significantly reducing the signaling overhead within the AirComp system assisted by a large-size RIS.
In addition, the authors in \cite{li2022reconfigurable} present an RIS-assisted multi-cell AirComp, where the RIS is leveraged to balance the intra-cell interference alignment and inter-cell interference mitigation, as well as overcoming unfavorable propagation environment.

\subsection{UAV-Aided Network}

\begin{figure}[t]
	\centering
	\includegraphics[scale=0.55]{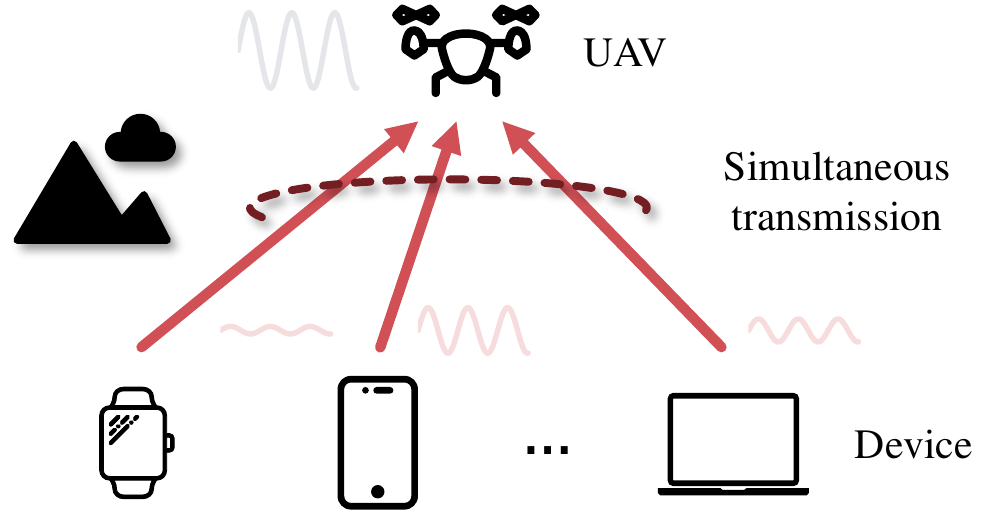}
	\caption{AirComp over a UAV-aided network.}
	\label{fig:uav}
\end{figure}

In the above studies on AirComp, the devices are assumed to be static or can only move within the area covered by static ground FCs.
However, large amounts of data are generally generated on mobile devices in practice, e.g., smart phones and vehicles, which can move over a wide area and even move out of the service area.
Meanwhile, the infrastructure for wired/wireless communication networks can be unavailable due to unpredictable natural disasters and other catastrophic situations.
Such scenarios prevent us from achieving efficient AirComp and thus a flexible communication system is required to tackle these issues.
Fortunately, UAV is emerging as a promising cost-effective and flexible equipment to assist the terrestrial communication networks, which can establish \textit{line-of-sight} (LoS) air-ground channels at a high altitude and promote the communication performance through the mobility control \cite{wu2021comprehensive}.
Therefore, the UAV-mounted FC can provide flexible network services for devices distributed in complicated environments, which has been exploited in \cite{farajzadeh2020mobility, joung2021overtheair, fu2021uav, zeng2022optimized, fu2023uav, jung2022performance} to facilitate AirComp, as shown in Fig. \ref{fig:uav}.

Specifically, the authors in \cite{farajzadeh2020mobility} leverage the mobility of UAV to reduce the computation error of AirComp in backscatter sensor networks, where the UAV serves as both a power emitter and reader for efficient data aggregation from backscatter nodes.
Besides, the authors in \cite{joung2021overtheair} develop a UAV-empowered AirComp strategy with space-time line code, which enables a data-collection UAV to efficiently aggregate data from multiple sensing UAVs by fully exploiting the spatial diversity gain.
In addition, the authors in \cite{fu2021uav} propose a joint trajectory and transceiver design for UAV-aided AirComp to minimize the time-average MSE, where the devices continuously move within the considered task duration.
The simulation results in \cite{fu2021uav} imply that the UAV tends to track the mobility of sensors and approach the cluster with low power budget of sensors to balance the communication distance and sensor transmit power of multiple clusters, thereby increasing the average computation accuracy of AirComp during multiple slots.
Instead of tacking mobile devices, the UAV employed in \cite{zeng2022optimized} is to serve the sensors distributed over a large area, where the UAV aggregates the data via AirComp from a dedicated subgroup of sensors within each time slot.
To promote the aggregation efficiency, the authors in \cite{zeng2022optimized} propose a $k$-means algorithm \cite{kanungo2002efficient} based grouping method to minimize the sum distance from the grouped sensors to the UAV, along with the shortest path scheme based on ant colony optimization \cite{dorigo2006ant} for UAV trajectory manipulation, which effectively reduces the computation MSE of a large-scale data aggregation.
Moreover, multiple UAVs are exploited in \cite{fu2023uav} to provide flexible AirComp services for multi-cluster networks, which demonstrates a trade-off between the number of completed data aggregations and the access delay of each device.
Moreover, the performance of UAV-enabled AirComp is theoretically analyzed in \cite{jung2022performance} to capture the impact of channel estimation and synchronization errors on the MSE, which implies that the imperfect CSI and synchronization induce an MSE error floor regardless of the SNR.

\subsection{Summary and Challenges}

In a nutshell, single-cell AirComp as the basic scenario focuses on exploring underlying techniques to harness intra-cell interference for efficient and accurate function computation, where digital linear-structured codes and analog transceiver design are employed for enabling coded and uncoded AirComp, respectively.
As for multi-cell networks, interference management strategies, including orthogonal resource allocation, SIA precoding, and cooperative power control, are developed to support multiple AirComp tasks in different cells.
Hierarchical AirComp leverages a multi-hop topology to handle the large-scale data aggregation with the studies on relay protocols and resource allocation.
In decentralized scenarios, AirComp accelerates the data consensus among different devices by allowing simultaneous neighboring data collection and update within single time slot.
Moreover, RIS-aided networks enhance the accuracy of AirComp through additional reflective beamforming design, while UAV-aided networks introduce mobility to empower flexible AirComp with trajectory optimization and multi-UAV coordination.
There are still several challenges to be addressed as follows:
\begin{itemize}
	\item \textbf{Coexistence Services:}
	Instead of focusing on stand-alone AirComp, the co-existence of AirComp and traditional communication services needs to be considered in practical scenarios \cite{du2021interference, ni2022integrating}.
	In such systems, only partial intra-cell interference can be harnessed to compute the desired function via AirComp and other interference should be reduced while satisfying the transmission requirements of other devices, which underscores the importance of interference management for such situations.
	
	\item \textbf{Error Amplification:}
	In hierarchical networks, the communication error induced at each hop may be amplified by the AF relay in the next hop, potentially leading to worse computation results in multi-hop AirComp compared to single-hop scenarios \cite{lin2022relayassisted}. 
	Hence, further investigations are needed to analyze the impact of error amplification on the computation accuracy of hierarchical AirComp.
\end{itemize}


\section{Information Theory Perspective} \label{sec:information_theory_tools}

Instead of quantifying the system performance in terms of the transmission rate to measure the maximum number of bits that can be successfully delivered per second, AirComp utilizes the computation rate to specify the maximum number of functions that can be reliably computed, which can be regarded as a reinterpretation of information theory from a function computation perspective.
In this section, we will review the existing works that analyze and maximize the computation rate for coded and uncoded AirComp from the information theory perspective.
A summary of representative works can be previewed in Table \ref{tab:it}.

\begin{table*}[t]
	\centering
	\caption{Representative Works on AirComp from the Information Theory Perspective}
	\label{tab:it}
	\scriptsize
	\begin{threeparttable}
	\renewcommand{\arraystretch}{1.5}
	\resizebox{\linewidth}{!}{
	\begin{tabular}{!{\vrule width1pt}c!{\vrule width1pt}m{0.27\linewidth}|c|>{\centering\arraybackslash}m{0.2\linewidth}|m{0.31\linewidth}!{\vrule width1pt}}
		\Xhline{1pt}
		\rowcolor[HTML]{e8e8e8}
		\bf{Category} & \multicolumn{1}{c|}{\bf{Workflow}} & \bf{Reference} & \bf{Channel model} & \multicolumn{1}{c!{\vrule width1pt}}{\bf{Feature}} \\ \Xhline{1pt}
		\multirow{10}{*}{\shortstack{Coded \\ AirComp}} & \multirow{10}{*}{\begin{tabular}
				{@{}m{\linewidth}@{}}\textbf{1. Device $k \in \{1, 2, \dots, K\}$:} \\ 1) Pre-processing: $\varphi_k(\cdot): \mathbb{R}^T \rightarrow \mathbb{R}^T$ \\ 2) Quantization: $\mathcal{Q}(\cdot): \mathbb{R}^T \rightarrow \{0, 1\}^{bT}$  \\ 3) Source encoder: $\mathcal{E}_{{\rm sc}, k}(\cdot): \{0, 1\}^{bT} \rightarrow \mathbb{Z}_p^m$ \\ 4) Channel encoder: $\mathcal{E}_{{\rm ch}, k}(\cdot): \mathbb{Z}_p^m \rightarrow \mathbb{C}^L$  \\ \textbf{2. Superposition over MAC} \\ \textbf{3. FC:} \\ 1) Channel decoder: $\mathcal{D}_{\rm ch}(\cdot): \mathbb{C}^L \rightarrow \mathbb{Z}_p^m$ \\ 2) Source decoder: $\mathcal{D}_{\rm sc}(\cdot): \mathbb{Z}_p^m \rightarrow \mathbb{R}^T$ \\ 3) Post-processing: $\psi(\cdot): \mathbb{R}^T \rightarrow \mathbb{R}^T$
		\end{tabular}} \raggedright & \cite{nazer2007computation} & Discrete linear and Gaussian MACs & Pioneer introduction and definition of computation rate. \\
		& & \cellcolor[HTML]{f4f4f2} \cite{goldenbaum2015nomographic} & Constant MACs \cellcolor[HTML]{f4f4f2} & Computation over a multi-cluster network. \cellcolor[HTML]{f4f4f2} \\
		& & \cite{nazer2011compute} & Gaussian MACs & Proposal of ``compute-and-forward'' relaying scheme. \\
		& & \cellcolor[HTML]{f4f4f2} \cite{jeon2016opportunistic} & Fading MACs \cellcolor[HTML]{f4f4f2} & Opportunistic transmission at devices. \cellcolor[HTML]{f4f4f2} \\
		& & \cite{chen2019communicating} & Uniform and non-uniform MACs & Comparison between coded AirComp and conventional OMA scheme. \\
		& & \cellcolor[HTML]{f4f4f2} \cite{chen2021toward} & Fading MACs \cellcolor[HTML]{f4f4f2} & CSI acquisition with \textit{automatic repeat request} (ARQ) strategy. \cellcolor[HTML]{f4f4f2} \\
		& & \cite{wu2019computation} & Fading MACs & Wideband transmission with OFDM. \\
		& & \cellcolor[HTML]{f4f4f2} \cite{wu2020nomaenhanced} & Fading MACs \cellcolor[HTML]{f4f4f2} & Computation assisted by power-domain NOMA. \cellcolor[HTML]{f4f4f2} \\
		& & \cite{chen2020computation} & Fading MACs & Computation over a multi-layer network. \\ \hline
		\multirow{6}{*}{\shortstack{Uncoded \\ AirComp}} & \multirow{6}{*}{\begin{tabular}
				{@{}m{\linewidth}@{}}\textbf{1. Device $k \in \{1, 2, \dots, K\}$:} \\ 1) Pre-processing: $\varphi_k(\cdot): \mathbb{R}^T \rightarrow \mathbb{R}^T$ \\ 2) Encoder: $\mathcal{E}_k(\cdot): \mathbb{R}^T \rightarrow \mathbb{C}^L$ \\ \textbf{2. Superposition over MAC} \\ \textbf{3. FC:} \\ 1) Decoder: $\mathcal{D}(\cdot): \mathbb{C}^L \rightarrow \mathbb{R}^T$ \\ 2) Post-processing: $\psi(\cdot): \mathbb{R}^T \rightarrow \mathbb{R}^T$
		\end{tabular}} \raggedright & \cellcolor[HTML]{f4f4f2} \multirow{4}{*}{\cite{nazer2007computation}} & \cellcolor[HTML]{f4f4f2} \multirow{4}{*}{Arbitrary MACs} & \cellcolor[HTML]{f4f4f2} \multirow{4}{*}{/} \\[11mm]
		& & \multirow{4}{*}{\cite{goldenbaum2013robust, goldenbaum2015onachievable}} & \multirow{4}{*}{Fading MACs} & \multirow{4}{*}{Coarse frame synchronization requirement.} \\[11mm]
		\Xhline{1pt} 
	\end{tabular}
	}
	
	\begin{tablenotes}
		\scriptsize
		\item $b$: number of quantization bits, $m$: natural number less than or equal to $T$, $p$: prime number larger than or equal to $(K (2^b - 1) + 1)^{T / m}$ \cite{goldenbaum2015nomographic}
	\end{tablenotes}
	\end{threeparttable}
\end{table*}

\subsection{Coded AirComp}

For coded AirComp, each device can encode its data into linear-structured codes, e.g., nested lattice code.
Since the integer combination of linear-structured codes can still fall into the field of itself, the same decoder used for decoding individual data can also be leveraged to recover superimposed signals, which makes coded AirComp easily to be implemented in existing digital wireless communication systems.

Recently, several works have analyzed the computation rate of coded AirComp under different system settings.
Specifically, the authors in \cite{nazer2007computation} present the computation rates of coded AirComp over discrete linear and Gaussian MACs with joint source-channel computation coding, which outperform the scheme with separated source-channel coding in terms of the computation rate.
The authors in \cite{nazer2011compute} apply coded AirComp to the ``compute-and-forward'' relaying strategy, where the relays first obtain the linear functions of transmitted signals via coded AirComp and then send these function values to the receiver.
Herein, the computation rate is analyzed based on the average probability of achieving a target computation accuracy at all relays, and then the coefficients for combining transmitted messages are designed to maximize the computation rate over \textit{additive white Gaussian noise} (AWGN) channels \cite{nazer2011compute}.
In addition, the authors in \cite{goldenbaum2015nomographic} investigate the computation rate of coded AirComp over a multi-cluster network with constant channel gains, which demonstrates that the computation rate scales down by the total number of clusters if each cluster is activated in a time-division manner.

Instead of assuming ideal MACs as in the above works, the authors in \cite{jeon2016opportunistic} study the coded AirComp over fading MACs, where an opportunistic in-network computation framework is proposed to ensure that the computation rate is not limited by deep-fading devices.
Then, a long-term ergodic computation rate is analyzed in \cite{jeon2016opportunistic}, which shows that a multi-user diversity gain can be achieved by scheduling the devices with large channel gains to participate in the aggregation.
Besides, the authors in \cite{chen2019communicating} analyze and compare the achievable computation rates of coded AirComp and conventional OMA scheme, which indicates that coded AirComp is not always superior to the OMA scheme under different settings in terms of the function computation.
Furthermore, the authors in \cite{wu2019computation} extend AirComp from narrowband to wideband fading channels, where each subcarrier is exploited to compute a sub-function via coded AirComp, and then the desired function is reconstructed from the received sub-functions.
The computation rate of such a wideband scenario is then derived in \cite{wu2019computation}, followed by an optimal power control for improving the computation rate while achieving a non-vanishing property.
Moreover, the authors in \cite{wu2020nomaenhanced} leverage the power-domain NOMA to enable multiple sub-functions to be computed over non-orthogonal resource blocks in a wideband scenario, which increases the computation rate and spectral efficiency.
The outage probability and diversity order are developed in \cite{wu2020nomaenhanced}, which reveal that the device with the worst channel gain imposes a performance limitation as the power goes to infinity.
Such a function division method is also utilized in a multi-hop network with a computation rate analysis \cite{wu2021computation}.
In addition, the transceiver design is taken into account to maximize the achievable computation rate of hierarchical AirComp \cite{chen2020computation} and to balance the trade-off between the computation rate and the transmission delay for CSI acquisition \cite{chen2021toward} .

\subsection{Uncoded AirComp}

Uncoded AirComp is able to simply map each pre-processed data to one channel input symbol with linear power scaling, which leads to the number of channel uses, $L$, being equal to the volume of generated data, $T$.
Hence, the computation rate of uncoded AirComp can reach one, i.e., $R_{\rm C} = 1$, according to Definition \ref{def:comp_rate} over the ideal MACs that are free of fading and noise \cite{goldenbaum2015onachievable}.
Beyond that, the authors in \cite{nazer2007computation} analyze the computation rate of uncoded transmission for sending arbitrary function over arbitrary MACs, which shows that the corresponding lower bound for the \textit{minimum mean-squared error} (MMSE) in function estimation exponentially increases with an attenuation factor of $1 / R_{\rm C}$.
In addition, the authors in \cite{goldenbaum2013robust} propose a robust scheme for function computation with uncoded transmission, where each device encodes its data into a series of power-scaled random signal sequences and the receiver is equipped with an energy detector to obtain the sum of transmit energy, such that only a coarse block synchronization is required.
The computation rate of such a robust analog transmission design is given in \cite{goldenbaum2015onachievable} to capture the trade-off between the efficiency and reliability of the function computation, which is determined by the function to be computed and the desired level of accuracy.

\subsection{Summary and Challenges}

In short, coded AirComp uses linear-structured codes for data encoding, where the decoder for individual data can also be used to tackle superimposed signals.
Besides, uncoded AirComp simplifies data mapping to channel input symbols with linear power scaling, which is able to achieve a computation rate of one over ideal MACs.
There are several challenges remain to be addressed as follows:
\begin{itemize}
	\item \textbf{Computation Rates Under Different Coding Strategies:}
	The existing computation rates of coded AirComp are mainly analyzed based on the nested lattice coding strategy, while the research and analyses of other feasible coding strategies remain open issues to be explored.
	
	\item \textbf{Compatible Coding Strategies:}
	Current wireless communication systems mainly adopt digital modulations, making it worthwhile to explore how to customize coded AirComp to be compatible with widely used modulation schemes to follow the evolution of wireless networks.
\end{itemize}


\section{Signal Processing Perspective} \label{sec:signal_processing_tools}

To achieve a desired magnitude alignment of superimposed signals, it is crucial to jointly design the transmitters and the receiver to compensate for the non-uniform channel fading between the FC and each of the devices, which stimulates the development of transceiver design with various advanced signal processing techniques.
In this section, we shall introduce the transceiver design for AirComp systems under different wireless communication scenarios from a signal processing perspective.
A summary of representative works reviewed in this section is provided in Table \ref{tab:sp}.

\begin{table*}[t]
	\centering
	\caption{Representative Works on AirComp from the Signal Processing Perspective}
	\label{tab:sp}
	\scriptsize
	\renewcommand{\arraystretch}{1.5}
	
	\begin{threeparttable}
	\resizebox{\linewidth}{!}{
	\begin{tabular}{!{\vrule width1pt}c!{\vrule width1pt}c|m{0.08\linewidth}<{\centering}|m{0.35\linewidth}|m{0.18\linewidth}<{\centering}|m{0.18\linewidth}!{\vrule width1pt}}
		\Xhline{1pt}
		\rowcolor[HTML]{e8e8e8}
		\multicolumn{1}{!{\vrule width1pt}c!{\vrule width1pt}}{\bf{Antenna}} & \multicolumn{1}{c|}{\bf{Reference}} & \multicolumn{1}{c|}{\bf{Objective}} & \multicolumn{1}{c|}{\bf{Transceiver design}} & \multicolumn{1}{c|}{\bf{Complexity}} & \multicolumn{1}{c!{\vrule width1pt}}{\bf{Feature}} \\ \Xhline{1pt}
		\multirow{12}{*}{SISO} & \cite{liu2020overtheair, cao2020optimized} & Instantaneous MSE and time-average MSE & $\bullet$ Threshold-based structure for static MACs \newline $\bullet$ Channel-inversion structure for time-varying MACs & $\mathcal{O}\left(K\right)$ & Optimal design with peak-power constraint. \\
		& \cellcolor[HTML]{f4f4f2} \cite{zang2020overtheair} & Instantaneous MSE \cellcolor[HTML]{f4f4f2} & $\bullet$ Equivalent convex reformulation \cellcolor[HTML]{f4f4f2} & \cellcolor[HTML]{f4f4f2} $\mathcal{O}\left(K\right)$ & Optimal design with sum-power constraint. \cellcolor[HTML]{f4f4f2} \\
		& \cite{qin2021overtheair} & Instantaneous MSE & $\bullet$ {\bf \textit{Transmitter design} (Tx):} KKT conditions \newline $\bullet$ {\bf \textit{Receiver design} (Rx):} First-order optimality condition & $\mathcal{O}\left(S^2 K + S^3\right)$ & Wideband transmission. \\
		& \cellcolor[HTML]{f4f4f2} \cite{frey2021corrchannel} & Confidence level with given MSE requirement \cellcolor[HTML]{f4f4f2} & $\bullet$ {\bf Tx:} Normalized transmit power with random phase shift \newline $\bullet$ {\bf Rx:} Energy detection of received signals across multiple channel uses while subtracting the noise energy based on the statistical information \cellcolor[HTML]{f4f4f2} & \cellcolor[HTML]{f4f4f2} $\mathcal{O}\left(L\right)$ & Computation over correlated MACs. \cellcolor[HTML]{f4f4f2} \\
		& \cite{liu2021spatialtemporal} & Instantaneous MSE & $\bullet$ {\bf Tx:} Equivalent convex reformulation \newline $\bullet$ {\bf Rx:} Kalman filter / Linear filter obtained via the first-order optimality condition & \shortstack{Kalman: $\mathcal{O}\left(K^3\right)$ \\ Linear: $\mathcal{O}\left(\ell\right)$} & Computation of spatial-and-temporal correlated signals. \\
		& \cellcolor[HTML]{f4f4f2} \cite{tang2020multi} & MSE \cellcolor[HTML]{f4f4f2} & $\bullet$ {\bf Tx:} Channel-inversion policy \newline $\bullet$ {\bf Rx:} First-order optimality condition \cellcolor[HTML]{f4f4f2} & \cellcolor[HTML]{f4f4f2} $\mathcal{O}\left(K\right)$ & Multi-slot aggregation with only once transmission per device. \cellcolor[HTML]{f4f4f2} \\ \hline
		\multirow{8}{*}{SIMO} & \cite{liu2020overtheair} & Instantaneous MSE & $\bullet$ Threshold-based structure for obtaining transmit scalars and the magnitude of receive beamforming vector \newline $\bullet$ Random generation for approximating the direction of receive beamforming vector & $\mathcal{O}\left(K + N_{\rm r}\right)$ & Near-optimal design. \\
		& \cellcolor[HTML]{f4f4f2} \cite{chen2018uniform} & Instantaneous MSE \cellcolor[HTML]{f4f4f2} & $\bullet$ {\bf Tx:} Uniform-forcing design \newline $\bullet$ {\bf Rx:} \textit{Semidefinite relaxation} (SDR) / \textit{Successive convex approximation} (SCA) \cellcolor[HTML]{f4f4f2} & \cellcolor[HTML]{f4f4f2} \shortstack{SDR: $\mathcal{O}\left(\max(K, N_{\rm r})^4 N_{\rm r}^{0.5}\right)$ \\ SCA: $\mathcal{O}\left(N_{\rm r}^3\right)$} & Proposal of uniform-forcing transceiver design. \cellcolor[HTML]{f4f4f2} \\
		& \cite{jiang2019overtheair} & Instantaneous MSE & $\bullet$ {\bf Tx:} Uniform-forcing design \newline $\bullet$ {\bf Rx:} \textit{Difference-of-convex} (DC) programming & $\mathcal{O}\left(I \max(K, N_{\rm r})^4 N_{\rm r}^{0.5}\right)$ & Computation assisted by RIS. \\
		& \cellcolor[HTML]{f4f4f2} \cite{fang2021overtheair} & Instantaneous MSE \cellcolor[HTML]{f4f4f2} & $\bullet$ {\bf Tx:} Uniform-forcing design. \newline $\bullet$ {\bf Rx:} SCA and Mirror-Prox algorithm. \cellcolor[HTML]{f4f4f2} & \cellcolor[HTML]{f4f4f2} \shortstack{$\mathcal{O}\left(K N_{\rm r} \log(N_{\rm r})\right)$} & Computation assisted by RIS. \cellcolor[HTML]{f4f4f2} \\
		& \cite{fang2021optimal} & Instantaneous MSE & $\bullet$ {\bf Tx:} Uniform-forcing design \newline $\bullet$ {\bf Rx:} \textit{Branch-and-bound} (BnB) algorithm & \shortstack{$\mathcal{O}\left(C^K N_{\rm r}^3 K^{3.5}\right)$ \\ $C = \frac{2 \pi}{\arccos\left(1 / \sqrt{1 + \epsilon_{\rm s}}\right)}$} & Optimal design for receive beamforming under the uniform-forcing optimization framework. \\ \hline
		\multirow{14}{*}{MIMO} & \cellcolor[HTML]{f4f4f2} \cite{zhu2019mimo} & Instantaneous MSE \cellcolor[HTML]{f4f4f2} & $\bullet$ {\bf Tx:} Uniform-forcing design \newline $\bullet$ {\bf Rx:} Differential geometry \cellcolor[HTML]{f4f4f2} & \cellcolor[HTML]{f4f4f2} $\mathcal{O}\left(K\max\{N_{\rm r}, N_{\rm t}\}^3\right)$ & Channel feedback via AirComp. \cellcolor[HTML]{f4f4f2} \\
		& \cite{chen2018overtheair} & Instantaneous MSE & $\bullet$ {\bf Tx:} Uniform-forcing design \newline $\bullet$ {\bf Rx:} Scaled unitary matrix & $\mathcal{O}\left(\left\lfloor \frac{N_{\rm r}}{Q}\right\rfloor K \max\{N_{\rm r}, N_{\rm t}\}^3\right)$ & Computation with antenna selection. \\
		& \cellcolor[HTML]{f4f4f2} \cite{wen2019reduced} & Instantaneous MSE \cellcolor[HTML]{f4f4f2} & $\bullet$ {\bf Tx:} Uniform-forcing design \newline $\bullet$ {\bf Rx:} Decomposed structure with inner and outer components \cellcolor[HTML]{f4f4f2} & \cellcolor[HTML]{f4f4f2} $\mathcal{O}\left(G\left(N_{\rm r}^3 + N_{\rm t} N_{\rm r} R\right)\right)$ & Computation over clustered IoT networks. \cellcolor[HTML]{f4f4f2} \\
		& \cite{zhai2021hybrid} & Instantaneous MSE & $\bullet$ {\bf Tx:} KKT condition \newline $\bullet$ {\bf Rx:} SCA / \textit{Block coordinate descent} (BCD) for receive analog beamforming, and the first-order optimality condition for receive digital beamforming & \shortstack{$\mathcal{O}\left(K N_{\rm t} N_{\rm r} N_{\rm rf}\right.$ \\ $\left.+ N_{\rm r}^2 K (Q + N_{\rm rf}) + C_{\rm hyb}\right)$ \\ SCA: $C_{\rm hyb} = N_{\rm r}^2(N_{\rm t} + N_{\rm rf})$ \\ BCD: $C_{\rm hyb} = N_{\rm r}^2 N_{\rm rf}^2$} & Hybrid beamforming design. \\
		& \cellcolor[HTML]{f4f4f2} \cite{hu2022ris} & Instantaneous MSE \cellcolor[HTML]{f4f4f2} & $\bullet$ {\bf Tx:} KKT condition \newline $\bullet$ {\bf Rx:} \textit{Riemannian conjugate gradient} (RCG) algorithm for receive analog beamforming and the first-order optimality condition for receive digital beamforming \cellcolor[HTML]{f4f4f2} & \cellcolor[HTML]{f4f4f2} \shortstack{$\mathcal{O}\left(K N_{\rm t} N_{\rm r} N_{\rm rf}\right.$ \\ $+ N_{\rm r}^2 K (Q + N_{\rm rf})$ \\ $\left. + K N_{\rm r}^2 N_{\rm rf}\right)$} & Computation assisted by RIS. \cellcolor[HTML]{f4f4f2} \\ 
		& \cite{chen2020highmobility} & Time-average MSE & $\bullet$ {\bf Tx:} SCA \newline $\bullet$ {\bf Rx:} SCA for receive analog beamforming and the first-order optimality condition for receive digital beamforming & \shortstack{$\mathcal{O}\left(K N_{\rm rf}^3 + \frac{K N_{\rm t} N_{\rm r} N_{\rm rf}}{\tau}\right)$} & Two-timescale hybrid beamforming design. \\
		\Xhline{1pt}
	\end{tabular}
	}

	\begin{tablenotes}
		\scriptsize
		\item $S$: number of subchannels, $\ell$: number of the most recent received signals, $N_{\rm r}$: number of receive antennas, $N_{\rm t}$: number of transmit antennas, $I$: number of iterations, 
		\item $\epsilon_{\rm s}$: solution precision, $Q$: data dimension, $G$: number of clusters, $R$: rank of cluster's covariance matrix, $N_{\rm rf}$: number of \textit{radio-frequency} (RF) chains, $\tau$: long-term slots
	\end{tablenotes}
	\end{threeparttable}
\end{table*}

\subsection{SISO AirComp} 

The transceiver design in SISO AirComp is to find appropriate transmit and receive scaling factors in one-dimensional complex scalar field, which is more tractable than the multi-dimensional optimization in multi-antenna scenarios.
To further simplify the transceiver design, the authors in \cite{liu2020overtheair} demonstrate that the phase of transmit scaling factor at each device can be designed to compensate for the phase induced by the complex receive scaling factor and the known channel coefficient, while its magnitude does not need to be changed and is merely constrained by the transmit power budget.
Hence, the transceiver design in SISO AirComp can be reduced to the joint design of transmit power at the devices and the receive scaling factor at the FC, where the optimization is performed in the real number field.

By assuming that the perfect CSI is available and all devices are well synchronized, the optimal transceiver design under different system constraints has been studied on SISO AirComp systems \cite{liu2020overtheair, cao2020optimized, zang2020overtheair}.
Considering the peak-power constraint at each device, the optimal transmit power of each device follows a threshold structure, where the threshold is determined by the receive scaling factor (also known as the denoising factor) and the quality indicator that accounts for both the channel conditions and power constraints at different devices \cite{liu2020overtheair, cao2020optimized}.
It suggests that the device can utilize the channel-inversion policy to perfectly compensate for the channel fading if its quality indicator exceeds the threshold, otherwise the full-power transmission should be applied.
Besides, a larger noise variance yields a higher threshold and asks for more devices to adopt full-power transmission, which requires the receiver to implement a smaller receive scaling factor to suppress the error caused by channel noise \cite{cao2020optimized}.
Instead of considering static channel conditions, the time-varying fading channel, a more general case, is also studied in \cite{liu2020overtheair, cao2020optimized}.
In particular, an optimal transceiver design of the time-varying case is proposed in \cite{cao2020optimized}, where the power control is shown to have a regularized channel-inversion structure for balancing the trade-off between the signal-magnitude alignment and the noise suppression.
Besides, the authors in \cite{liu2020overtheair} investigate the ergodic performance of SISO AirComp in terms of the time-averaging MSE and power consumption, which reveals the trade-off between the computation effectiveness and the energy efficiency.
In addition to the peak-power constraints, the authors in \cite{zang2020overtheair} propose an optimal design for minimizing the MSE under the sum-power constraint, where the solution can be obtained in a closed form from the equivalent convex problem.
Besides, the optimal design for sum-power minimization under the MSE requirement is also studied in \cite{zang2020overtheair}.

Unlike the above works that consider a single radio channel, a wideband AirComp system is studied in \cite{qin2021overtheair} to increase the computation accuracy. This is done by exploiting the multi-channel diversity, where each device broadcasts the signal over multiple selected subchannels under a transmit power constraint.
The results in \cite{qin2021overtheair} indicate that, in order to enhance the aggregation accuracy, the devices prefer to choose a subset of channels with good conditions for their data transmission, instead of simply broadcasting signals over all subchannels.
Besides, the analyses on MSE minimization within real and complex domains in \cite{qin2021overtheair} reveal that the optimization in real domain can separately process the real and imaginary parts of the signal and align them onto a one-dimensional signal space, which is able to yield a smaller MSE than the optimization in complex domain that uses a complex number to directly scale the complex distorted signal.
In addition, the authors in \cite{tang2020multi} propose a multi-slot AirComp to reduce the signal distortion by distributing the transmissions over multiple slots.
Similar to the broadband AirComp considered in \cite{qin2021overtheair}, multi-slot AirComp enables devices to select the slot with strong channel gain for data transmission, thereby escaping the deep fading that may be confronted with only single-slot transmission.

Different from the aforementioned works that assume independent channel fading and independent transmitted signals among different devices, the AirComp systems with correlated channels and correlated signals are studied in \cite{frey2021corrchannel} and \cite{liu2021spatialtemporal}, respectively.
In particular, the authors in \cite{frey2021corrchannel} propose a scheme for AirComp in a general setting of fast fading channels that can be non-Gaussian and correlated.
For correlated signals, the authors in \cite{cao2020optimized} demonstrate that the corresponding MSE minimization problem can be transformed into a tractable upper bound minimization problem, which reduces to the optimization with independent signals as proposed in \cite{liu2020overtheair, cao2020optimized}.
Furthermore, the authors in \cite{liu2021spatialtemporal} implement an optimal policy for AirComp with spatial-and-temporal correlated signals.
Herein, the transceiver design is achieved through the Kalman filter in an iterative prediction-correction way, which is also valid for spatial-and-temporal independent signals \cite{liu2021spatialtemporal}.
Also, a low-complexity design is proposed in \cite{liu2021spatialtemporal} by applying a linear filter to the stored historical signals as the current computation output, which achieves the same performance as the optimal design with a sufficient filter length.

\subsection{SIMO/MIMO AirComp}

To further enhance the performance of AirComp, multi-antenna technology is a powerful tool to increase the diversity order and degrees of freedom in the transceiver design \cite{chen2018uniform, chen2018overtheair, zhu2019mimo}.
Specifically, an FC with multiple antennas is able to increase the diversity gain for reducing the distortion of aggregated signals \cite{chen2018uniform}.
Besides, equipping multiple antennas at both the FC and the devices makes it possible to simultaneously compute multiple functions of multi-modal sensing data, where various data from multiple devices can be aggregated at different antennas \cite{chen2018overtheair, zhu2019mimo}.
In such cases, the receive/transmit scaling factor becomes a multi-dimensional complex beamforming vector, which complicates the formulation of MSE and leaves the optimal transceiver design to be solved for SIMO/MIMO AirComp.
Even though, a sub-optimal method, called uniform-forcing transceiver design, is developed in \cite{chen2018uniform} to facilitate the optimization on SIMO/MIMO AirComp.
The uniform-forcing design allows all devices to perfectly compensate the equivalent fading channels during the transmission while satisfying their transmit power constraints, where the equivalent fading channel is a combination of the receive beamforming vector/matrix and the fading channel vector/matrix \cite{chen2018uniform, chen2018overtheair, zhu2019mimo}.
Under this scheme, the received signal at the FC becomes an unbiased estimation of the sum of transmitted symbols \cite{liu2020overtheair}, which forces the error caused by the signal-magnitude misalignment to be zero but at the cost of elevating the noise-induced aggregation error \cite{cao2020optimized}.
The uniform-forcing transceiver design makes the MSE minimization problem more tractable in multi-antenna scenarios, which stimulates a growing body of research to explore the possible enhancement in the function computation via SIMO/MIMO AirComp.

\subsubsection{SIMO AirComp}

By implementing the uniform-forcing design for SIMO AirComp, the MSE minimization problem becomes a min-max optimization problem and can be further reduced to a \textit{quadratically constrained quadratic programming} (QCQP) problem, which, however, is non-convex and \textit{non-deterministic polynomial-time} (NP) hard \cite{chen2018uniform}.
Even though, it is found that the resulting QCQP problem has the same mathematical form as the downlink multicast beamforming design \cite{sidiropoulos2006transmit}, which establishes AirComp-multicasting duality similar to the uplink-downlink duality for multiuser MIMO communications \cite{jindal2004on, chen2018uniform, zhu2019mimo}.
As mentioned in \cite{zhu2019mimo}, the AirComp-multicasting duality holds because both the receive beamforming in AirComp and the multicast beamforming in multicasting aim to be aligned with multiple fading channels but with different targets, i.e., reducing MSE in AirComp and increasing minimum SINR in multicasting.
Therefore, based on such a duality, the existing optimization methods for multicasting problem can be directly leveraged to tackle the MSE minimization problem in SIMO AirComp, such as SDR \cite{sidiropoulos2006transmit}, SCA \cite{tran2014conic}, and BnB algorithms \cite{lu2017efficient}.

In particular, the non-convex QCQP problem for the MSE minimization can be converted into a \textit{semidefinite programming} (SDP) problem, which can be solved with the SDR technique by dropping the rank-one constraint \cite{luo2010semidefinite}.
Due to the weak capability of obtaining the rank-one solution of the SDR technique in high-dimensional space \cite{luo2007approximation}, the authors in \cite{chen2018uniform} further propose an SCA algorithm by iteratively approximating the non-convex quadratic constraints to convex linear constraints, which significantly enhances the solution quality initiated by the SDR algorithm.
Also to overcome the limitations of the SDR technique, the authors in \cite{jiang2019overtheair} propose a DC programming algorithm to induce the rank-one solution of the SDP problem.
The main idea of the DC algorithm is to represent the rank-one constraint as a DC function and then set it as a penalty term in the objective function.
By linearizing the concave term in the DC function, the reformulated problem can converges to a rank-one solution of the original SDP problem through multiple iterations \cite{tao1997convex}.
Motivated by \cite{lu2017efficient}, the authors in \cite{fang2021optimal} propose a globally optimal design for the resulting QCQP problem based on the BnB algorithm, where the solution is induced by iteratively reducing upper bound and lifting lower bound for the objective function until they converging to the same value with judiciously designed branching strategy.
In contrast to the above works that focus on solving the QCQP problem, the authors in \cite{fang2021overtheair} propose an SCA-based algorithm to tackle the original min-max problem.
As the subproblem in each SCA iteration can be transformed into a smooth convex-concave saddle point problem, the authors in \cite{fang2021overtheair} further utilize the Mirror-Prox method \cite{nemirovski2004prox} to solve each subproblem with a low computational complexity, which is much friendly to accommodate the scenarios with massive devices.
Moreover, the authors in \cite{liu2020overtheair} develop an algorithm to find near-optimal solutions for SIMO AirComp.
By denoting the receive beamforming vector as the product of a scaling factor and a unit vector, it is able to first find the optimal receive and transmit scaling factors by utilizing the optimal design developed for SISO AirComp \cite{cao2020optimized, liu2020overtheair}, and then find the suboptimal unit vector that induces the minimum MSE in a sequence of randomly generated unit vectors.

\subsubsection{MIMO AirComp}

The uniform-forcing design for MIMO AirComp enables the transmit beamforming matrices at devices to not only compensate for non-uniform channel fading, but also eliminate the interference caused by data for computing different functions \cite{chen2018overtheair}.
Besides, the receive beamforming matrix can be set as a scaled unitary matrix, which is a sub-optimal setting but effectively simplifies the receiver design with marginal performance loss in MIMO systems \cite{medra2015incremental, zhu2019mimo}.
To reduce the overhead for additional CSI acquisition at the FC, the authors in \cite{chen2018overtheair} propose an antenna selection scheme when the number of receive antennas is larger than the number of functions to be computed.
Instead of selecting a subset of receive antennas at the FC, the authors in \cite{zhu2019mimo} formulate an approximate problem for optimizing the receive beamforming matrix on a Grassmann manifold by tightening the power constraints, and then exploit the differential geometry method to induce a closed-form solution.

Although large-antenna arrays provide rich spatial degrees of freedom and tremendous beamforming gain, they inevitably lead to high complexity in the transceiver design for MIMO AirComp.
To tackle this issue, the authors in \cite{wen2019reduced} propose a reduced-dimension receive beamforming design in clustered IoT networks, where the channel model is characterized by the clustered transmitters and rich local scattering \cite{adhikary2013joint}.
By exploiting the structure of clustered MIMO channels, the optimal aggregation beamformer is decomposable with inner component for channel-dimension reduction and outer component for equalization of the low-dimensional small-scale fading channels \cite{wen2019reduced}.
Furthermore, the authors in \cite{zhai2021hybrid} propose a hybrid beamforming design for MIMO AirComp.
To minimize the MSE, an alternating-optimization-based algorithm is considered in \cite{zhai2021hybrid} by jointly optimizing the transmit digital beamforming at devices and the receive hybrid beamforming at the FC, where the transmit, receive analog, and receive digital beamformings are alternately obtained via the Lagrangian duality method, SCA or BCD method, and the first-order optimality condition, respectively.
With such a hybrid beamforming design, the required RF chains at the FC as well as the complexity of digital processing can be significantly reduced while achieving a close performance to the MIMO AirComp with a fully-digital receiver.
Besides, the authors in \cite{hu2022ris} propose a low-complexity RCG algorithm to obtain the receive analog beamforming in hybrid beamforming design for MIMO AirComp.
Moreover, the authors in \cite{chen2020highmobility} propose a mixed-timescale hybrid combining scheme to minimize the average MSE for multi-modal sensing via MIMO AirComp, where the baseband combiner at the FC is designed according to the real-time effective CSI, and the RF combiner at the FC and the transmit beamforming of devices are adapted to the long-term statistical CSI.

\subsection{Summary and Challenges}

To sum up, the transceiver design for SISO AirComp with peak/sum-power constraints, wideband communication, and correlated channels/signals have been explored.
For SIMO and MIMO AirComp, the uniform-forcing transceiver design simplifies the corresponding optimization.
Meanwhile, AirComp-multicasting duality facilitates the application of multicasting optimization methods to SIMO AirComp. 
Besides, hybrid beamforming, reduced-dimension beamforming, and antenna selection schemes address the high-complexity transceiver design due to large-antenna arrays in MIMO AirComp. 
To further enhance the performance of AirComp, the following challenges need to be addressed:
\begin{itemize}
	\item \textbf{Mobile Data Aggregation:}
	The existing transceiver design is mainly proposed for data aggregation with static devices and how to efficiently aggregate data from mobile devices via AirComp while mitigating the Doppler effect is challenging to be investigated.
	
	\item \textbf{Theoretical Performance Analyses:}
	In order to statistically quantify and compare the performances among various transceiver design methods in general scenarios, it is necessary to derive the corresponding theoretical performance in terms of computation accuracy of AirComp to guide practical implementations.
\end{itemize}

\section{Practical Implementation Issues} \label{sec:practical_implementation}

In this section, we will review the literature that considers system design for AirComp while taking into account the practical implementation issues.

\subsection{Efficient Channel Feedback}

In most of the existing studies on AirComp, perfect CSI is generally assumed to be available at the FC for implementing various advanced techniques, such as power control and beamforming design.
Since AirComp focuses on uplink transmission, time-division duplexing is commonly considered in the existing studies to achieve local CSI estimation at each device via the channel reciprocity based on the pilot signal broadcast by the FC.
However, the signaling overhead of the above training procedure is linearly scaling with the number of devices for CSI feedback, which may lead to high communication latency in ultra-dense wireless networks.
To address this issue, it is required to reduce the overhead for implementing system design without sacrificing performance.

Fortunately, various channel feedback approaches have been developed to realize efficient AirComp design while avoiding massive overhead for CSI gathering.
Specifically, the authors in \cite{chen2018cooperativewideband} and \cite{chen2018overtheair} exploit the ``OR'' property \cite{katti2008xors} of wireless channels to determine the optimal receive scaling factor at the FC, which is equal to the minimum ratio of the channel gain to the effective signal power among different devices.
Instead of obtaining global CSI at the AP, each device quantizes the corresponding ratio into a binary sequence based on its local CSI, and then the FC can obtain the binary representation of the minimum ratio by determining one significant bit per feedback \cite{chen2018cooperativewideband, chen2018overtheair}.
As the overhead of this method is mainly determined by the length of the quantized binary sequence, it is able to significantly reduce the time duration for CSI gathering as compared with the conventional approach that collects the global CSI \cite{chen2018overtheair}.
Besides, ARQ strategy is exploited in \cite{chen2021toward} to avoid massive CSI aggregation in AirComp systems, which effectively reduces the overhead by allowing multiple devices to concurrently transmit their data according to the threshold in each time slot without sequentially feeding back their own CSI to the FC \cite{chen2021toward}.
In addition to using AirComp for achieving efficient WDA, the authors in \cite{ang2019robust, chen2020computation, zhu2019mimo} also leverage AirComp to accelerate the system optimization, where the FC is able to directly obtain the optimal receive beamforming design by aggregating the feedback signals based on the local CSI via AirComp.
Furthermore, similar to the quantization method considered in \cite{chen2018cooperativewideband} and \cite{chen2018overtheair}, the authors in \cite{zhu2019mimo} propose an AirComp-based feedback for optimizing the receive scaling factor with several feedback rounds.
By assuming the channel reciprocity, the above method enables each device to obtain the local CSI and transmit beamforming design via downlink broadcasting, and allows the FC to optimize the receive beamforming via AirComp, which requires only a single symbol duration in each transmission phase.
Therefore, it is able to achieve efficient system design with a time complexity independent of the number of devices.

\subsection{Robust Design}

In practice, the transmitters and receivers can only obtain imperfect CSI due to various reasons, e.g., inaccurate channel estimation and finite-rate feedback \cite{love2008overview}.
Hence, simply implementing the system design derived from perfect CSI may inevitably cause performance degradation in practical wireless networks with imperfect CSI.
To address this issue, robust design is required to ensure a certain level of the system performance, which is generally characterized by two classes of models, namely stochastic (expectation-based) and deterministic (worst-case) models.
In the stochastic model, the uncertain CSI error follows a Gaussian distribution with known statistics (e.g., the mean and the covariance), based on which the system is designed to optimize the average or outage performance \cite{zhang2008statistically, gershman2010convex}.
In the deterministic model, the norm of uncertain CSI errors is bounded by a known spherical uncertainty region, which is employed to characterize the worst-case performance \cite{vorobyov2003robust, wang2009worstcase}.

As the CSI plays an important role in achieving the magnitude alignment at the FC, robust design is critical for AirComp to achieve a reliable function computation under imperfect CSI.
Specifically, the authors in \cite{huang2015robust} propose a robust transceiver design for MIMO AirComp with imperfect CSI modeled by the worst-case model.
The corresponding robust transceiver design is formulated as an optimization problem to minimize the worst-case MSE, which can be solved by an alternating optimization algorithm after converting the semi-infinite constraints into linear matrix inequalities.
To reduce the channel training complexity, the authors in \cite{ang2019robust} present an imperfect CSI model under over-the-air signaling procedure, where the channel uncertainty vector/matrix is directly added on the superimposed feedback of effective CSI.
The corresponding robust transceiver design is then proposed in \cite{ang2019robust} by considering both the stochastic and deterministic models, which achieves smaller MSE with a low training complexity than the methods assuming perfect CSI.
Furthermore, the authors in \cite{an2021robust} and \cite{zhang2022worstcase} propose the robust design for RIS-assisted AirComp, where the CSI of reflective links is imperfect due to the erroneous channel estimation and weak signal processing capability of the passive reflecting elements at the RIS.
Moreover, the transceiver design based on the statistics of channel estimation errors is developed in \cite{chen2022over} for SISO and SIMO AirComp, where a regularized channel inversion policy can be adopted to resist imperfect CSI in SISO case and an alternating optimization design based on the error statistics is necessary for the SIMO case.

Besides the perfect CSI acquisition, perfect synchronization is another ideal assumption in most of the existing studies on AirComp, which, however, is difficult to realize in large-scale networks.
If there are asynchronous concurrent signals arriving at the FC, the accuracy of estimated function may be severely deteriorated due to reversal superposition.
To address this issue, the authors in \cite{goldenbaum2013robust} propose a robust design with coarse synchronization by letting devices transmit random sequences at a transmit power being proportional to the real-valued sensor information.
Besides, the authors in \cite{shao2021federated} propose a maximum-likelihood estimation design for misaligned AirComp with residual channel-gain variation and symbol-timing asynchrony among devices.
As the maximum-likelihood estimation proposed in \cite{shao2021federated} is susceptible to noise, a Bayesian approach is further proposed in \cite{shao2021bayesian} by exploiting two pieces of statistical information of transmitted data, i.e., the first and the second sample moments, which addresses the error propagation and noise enhancement problems caused by the maximum-likelihood estimator.

\subsection{Blind Data Fusion}

As discussed above, most of the current design for AirComp relies on the instantaneous CSI at the FC and/or the devices.
Nonetheless, the cumbersome CSI acquisition inevitably introduces additional communication delay and signaling overhead, which stimulates the CSI-free design, also known as blind design, for AirComp systems.

Initially, to explore the impact of CSI on function computation in AirComp systems, the authors in \cite{goldenbaum2014on} conduct the performance analysis for the scenarios with different local CSI available at devices.
It shows that the channel magnitude information at devices is sufficient to achieve the same performance as the case with perfect local CSI, and no CSI is needed at devices if a multi-antenna FC has access to the statistical channel knowledge \cite{goldenbaum2014on}.
Then, the authors in \cite{dong2020blind} develop a blind AirComp to achieve desired function computation by leveraging the blind demixing, which is a powerful tool for recovering a sequence of signals without accessing any CSI \cite{ling2017blind}, and further propose a randomly initialized Wirtinger flow method to solve the resulting blind demixing problem with provable optimality guarantees.
To mitigate the possible destructive signal superposition due to the phase difference when CSI is not available, the authors in \cite{frey2021corrchannel} apply random phase shifts to the transmitted signals at different devices, and then average them over multiple channel uses to compensate for the phase difference and reduce the impact of additive noise.
In addition, the authors in \cite{sahin2021distributed} and \cite{sahin2021overtheair} leverage the one-bit quantization to achieve blind AirComp, where the receiver obtains the sign of aggregated signals based on majority vote by detecting the energy accumulated on different subcarriers \cite{sahin2021distributed} and superposed pulse-position modulation symbols \cite{sahin2021overtheair}, which eliminate the reliance on the CSI and relaxing the requirement for synchronization.
Furthermore, the balanced number system is employed in \cite{sahin2022over} to enable continuous-valued computations in the digital AirComp system without the need of CSI acquisition.

\subsection{Energy-Efficient and Sustainable Design}

The energy consumption for data transmission is a critical issue to be considered in sustainable IoT networks.
In order to prolong the network lifetime, the energy-efficient design is required for enhancing the applicability of AirComp in various scenarios.
For instance, the authors in \cite{basaran2020energy} propose an MMSE estimation scheme that achieves an energy-efficient AirComp by utilizing the spatial correlations among transmitted signals over AWGN channels.
Due to the spatial correlation, only a small number of observations are required to be sampled for the function computation while achieving an acceptable MSE level, which effectively reduces the energy consumption for data transmission and prolong the network life with densely deployed devices \cite{basaran2020energy}.
In addition, the authors in \cite{zhai2021power} propose a power minimization scheme for MIMO AirComp, where the transmit power is set as the optimization objective to be minimized under an MSE constraint.

Moreover, since IoT devices are usually energy-constrained and battery-powered, how to efficiently recharge the hard-to-replace batteries is another challenge to be considered to increase the sustainability of services provided by these devices.
Fortunately, the emergence of \textit{wireless power transfer} (WPT) technique enables the devices to harvest energy from the ambient RF signals radiated by the energy transmitter \cite{bi2015wireless}.
Hence, the integration of WPT and AirComp provides an attractive solution to achieve sustainable function computation.
Motivated by this, the authors in \cite{li2019wirelessly} develop a wirelessly powered AirComp framework by jointly optimizing the energy beamforming and transceiver design, where the FC serves as not only a power beacon but also a data FC.
Specifically, the FC first energizes the dispersed devices through the downlink energy beamforming and then the devices exploit the harvested energy for subsequent uplink AirComp, thereby ensuring a sustainable AirComp process \cite{li2019wirelessly}.
To further enhance the spectrum- and energy-efficiency in severe propagation environment, an RIS is leveraged in \cite{wang2021wirelesspowered} to assist the downlink WPT and uplink AirComp between the FC and devices, in which the deployment of the RIS not only reduces the downlink energy propagation loss but also improve the accuracy of the uplink function computation.

\subsection{Prototype Validation}

So as to put theory into practice, various prototypes have been built to validate feasibility and effectiveness of AirComp in real world.
Specifically, the authors in \cite{sigg2012calculation} propose to calculate mathematical functions via AirComp by encoding values into Poisson distributed burst sequences.
To verify the proposed transmission scheme, a wireless sensor network platform with $15$ simple nodes and a central receiver is developed in \cite{sigg2012calculation}, where each sensor node consists of one micro-controller (e.g., the Arduino Uno micro-controller board), one sensor, and a $2.4$ GHz sinusoid signal generator.
By implementing an experiment of mean temperature sensing, it is verified that the computation error can be reduced by increasing the length of burst sequences \cite{sigg2012calculation}.
The authors in \cite{kortke2014analog} establish a testbed to validate the transmission scheme proposed in \cite{goldenbaum2013robust}.
The implemented experiment is performed on self-developed multi-antenna software-defined radio devices, where one of these antennas is configured as the FC and others emulate separate single-antenna sensor nodes.
The corresponding experiment results demonstrate that the interference of MACs can indeed be usable for the computational function, despite some practical system impairments such as asynchronous transmission and additive noise.
Besides, the authors in \cite{chen2018cooperativewideband} and \cite{altun2017testbed} leverage the \textit{universal software radio peripheral} (USRP) platform to verify the proposed transmission schemes for AirComp.
The results obtained from the testbed established in \cite{altun2017testbed} reveal that the MSE increases in proportional to the input data value under imperfect channel estimations.
Moreover, the authors in \cite{guo2021overtheair} design a prototype based on the Xilinx software-defined radio platform to prove the concept of AirComp.
As the time and carrier frequency offsets in OFDM transmission induce phase noises to the transmitted signals, which may lead to the failure of coherent waveform superposition, an efficient protocol is proposed in \cite{guo2021overtheair} to address this issue by adding one more signaling round for phase error estimation and compensation.

\subsection{Summary and Challenges}

In summary, to tackle practical implementation issues, strategies like efficient channel feedback, robust design for imperfect CSI, and blind data fusion are developed to reduce the signaling overhead for efficient AirComp. 
Besides, energy-efficient and sustainable design contributes to green AirComp, and the prototype validation demonstrates the feasibility of AirComp in real-world scenarios.
There are still some challenges required to be tackled as follows:
\begin{itemize}
	\item \textbf{Receiver Amplitude Restriction:}
	As uncoded AirComp modulates the data to the signal magnitude, massive connectivity may lead to difficulties at receiver in accurately recovering superimposed signals with large amplitudes due to hardware limitation, which makes it challenging to achieve desired AirComp in practical systems.
	
	\item \textbf{Standardization:}
	Standardization is imperative for realizing widespread practical applications and commercial deployment.
	However, due to the paradigm shift from separate to integrated communication and computation, the standardization for AirComp needs to be further considered and discussed.
\end{itemize}


\section{Applications of AirComp} \label{sec:applications}

This section discusses the applications of AirComp in two categories of networks, i.e., IoT and EI networks, where IoT networks are introduced from the perspectives of distributed sensing and wireless control, while EI networks are emphasized by distributed learning and privacy preservation.

\begin{figure}[t]
	\centering
	\subfigure[Distributed sensing]{
		\centering
		\includegraphics[scale=0.62]{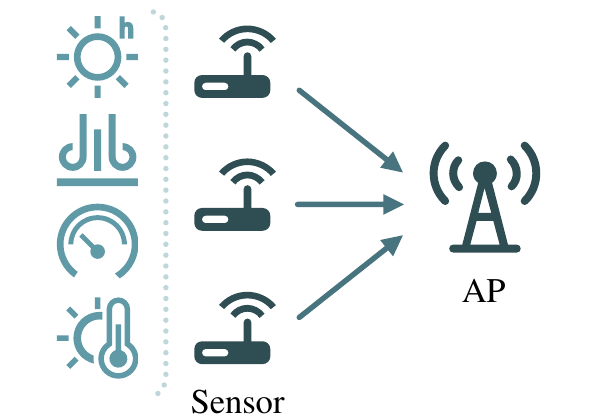}
		\label{fig:sensing}
	}
	\subfigure[Wireless control]{
		\centering
		\includegraphics[scale=0.62]{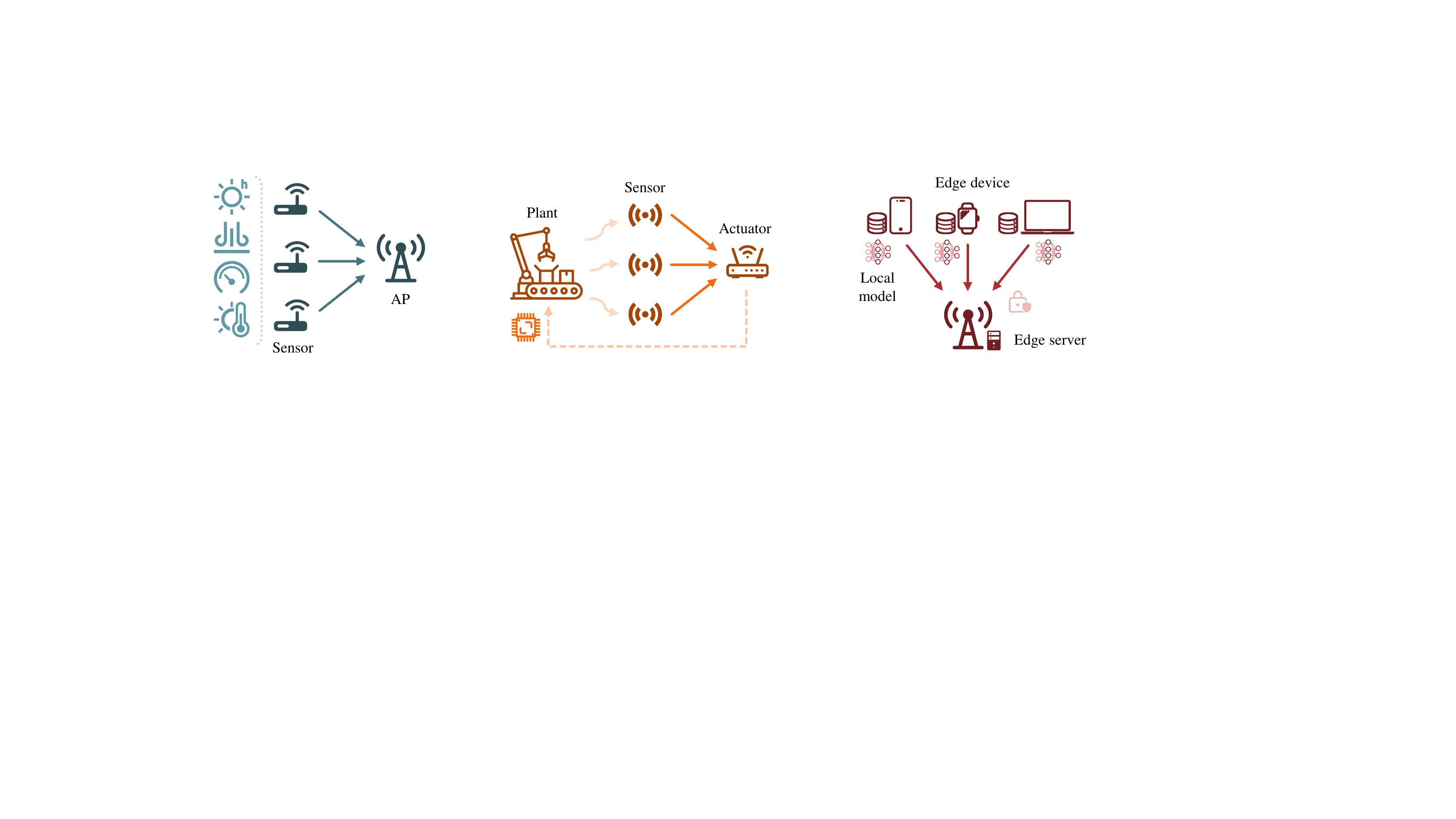}
		\label{fig:control}
	}
	\subfigure[Edge intelligence]{
		\centering
		\includegraphics[scale=0.62]{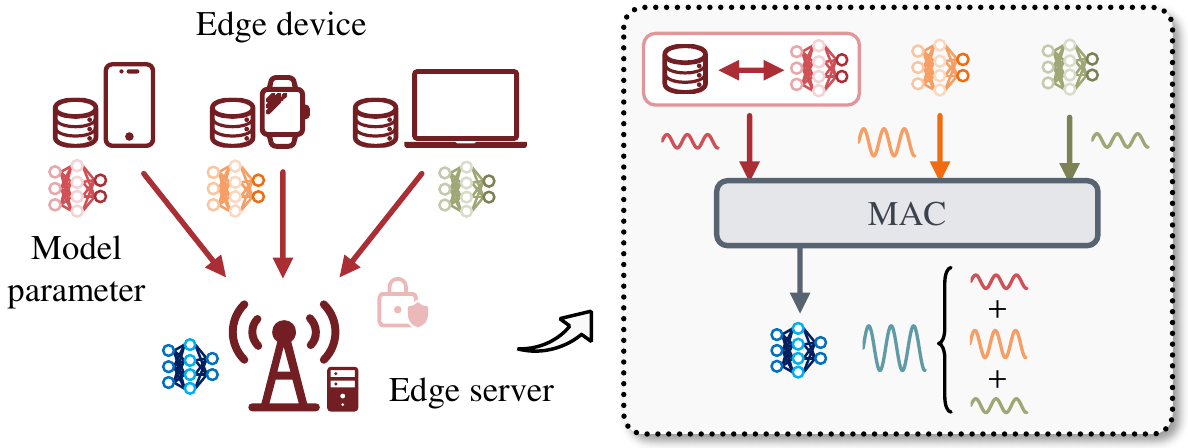}
		\label{fig:ei}
	}
	\caption{Application scenarios of AirComp.}
	\label{fig:app}
\end{figure}

\subsection{IoT Networks}

With the advancement of IoT techniques, pervasive devices are connected to the communication network for providing ubiquitous services, especially for distributed sensing and wireless control, where massive transient data needs to be aggregated in time for further processing and feedback.

\subsubsection{Distributed Sensing}

Distributed sensing aims to monitor and draw the digital picture of physical world with vastly and densely distributed sensors, e.g., the average temperature/humidity of an area for environmental monitoring, as shown in Fig. \ref{fig:sensing}, which are more interested in obtaining the function value of sensor readings rather than the individual raw data.
As such an objective is consistent with the paradigm of AirComp, it is able to achieve efficient distributed sensing by conducting simultaneous transmission among all sensors, while the desired function can be directly computed over the air.
Meanwhile, the characteristics of wireless sensor networks can be further leveraged to promote the performance of AirComp.
For instance, the channel correlation can be leveraged to facilitate the distributed sensing in the systems with relatively stable radio environment and locations of sensor nodes \cite{frey2021corrchannel}.
In densely deployed IoT networks, the adjacent devices shall generate spatial-and-temporal correlated observations due to the similarity of their settled environments, which can be used to reduce the aggregation error \cite{liu2021spatialtemporal} and the energy consumption \cite{basaran2020energy} of AirComp.
Benefiting from the WPT technique, wireless-powered AirComp can prolong the service time of devices for distributed sensing tasks \cite{li2019wirelessly, wang2021wirelesspowered}.
Moreover, UAV-empowered AirComp is capable of gathering the data from high-mobility and vast-distributed sensor nodes by constructing favorable channels \cite{zhu2019mimo, fu2021uav}, which augments the coverage and flexibility of the services.
In addition, the authors in \cite{tang2022radix} propose a radix-partition-based AirComp for remote state estimation in IoT networks, which can efficiently map the sensor measurements to the allocated resource pool without the activity detection.

\subsubsection{Wireless Control}

Wireless control system has been highly demanded in smart industry and agriculture recently, where a group of agents need to perform specific actions to accomplish an overall cooperative task by interacting with each other \cite{olfati2007consensus, cao2013overview, ayaz2019iot}, as shown in Fig. \ref{fig:control}.
Herein, each agent needs to iteratively gather information from others for updating its own state until convergence, which generally consists of two phases within each iteration, i.e., the communication phase for information exchanges and the computation phase for state updates.
By integrating these two phases, efficient network-wide consensus can be achieved via AirComp \cite{nazer2011local, zheng2012fast, goldenbaum2012nomographic, molinari2018exploiting, molinari2020exploiting, molinari2021maxconsensus}.
For instance, the authors in \cite{molinari2019efficient} exploit AirComp to support efficient wireless consensus for formation control, which effectively saves the radio resources for wireless control systems.
Besides, the authors in \cite{park2021wireless} employ AirComp to compute the control signals of wireless control systems, which constructs an over-the-air controller system with larger achievable stability region as compared with other wireless control systems. 
Furthermore, a power allocation strategy is proposed in \cite{park2022optimized} for AirComp-assisted closed-loop control system, where the actuator can efficiently aggregate the sensor measurements through AirComp to generate a controller feedback.
Then, by further combined with the control-optimal policy design developed in \cite{park2022optimized}, the control stability and transmit power consumption can be well balanced.
Moreover, the authors in \cite{kim2023control} apply the MIMO AirComp to achieve efficient aggregation of control signals, which is able to mitigate the negative effect of both the channel and measurement noises.

\subsection{EI Networks}

Edge intelligence aims to distill the intelligence from network edge by exploiting the geographically dispersed data and computing power \cite{shi2020communication, letaief2022edge}, as shown in Fig. \ref{fig:ei}, which calls for efficient coordination and communication between the server and edge devices.

\subsubsection{Distributed Learning}

Due to the data explosion in modern society, AI has been regarded as a revolutionary strategy to efficiently extract knowledge from voluminous data. 
In general, AI models are trained in a centralized server by collecting the datasets from dispersed devices, which, however, induces several drawbacks. 
For instance, the enormous data upload and re-storage in the server consume vast amount of time, energy, communication, and storage resources. 
Besides, data generated at local devices are usually privacy sensitive, which is violated when they are gathered to the central server for a centralized processing. 
To overcome these challenges, distributed learning techniques have been developed to enable dispersed devices to collaboratively train a shared global model efficiently. 
Chief among them is the renowned \textit{federated learning} (FL) framework \cite{mcmahan2017communication, yang2019federated, chen2021distributed, zhou2023toward}, where only model/gradient parameters are exchanged for ensuring the privacy of local data.

Despite avoiding the transmission of large volumes of data, limited radio resources is still a bottleneck for implementing FL in wireless networks, due to the high dimensional model/gradient parameters required to be periodically exchanged between the server and edge devices for complicated learning tasks. 
Therefore, conventional OMA schemes become inefficient for exchanging high-dimensional model parameters in terms of the communication delay by allocating orthogonal radio resources to participating devices.
As the objective of model aggregation is to obtain a weighted sum of local model updates that coincides with the nomographic function computation, AirComp can be employed to enable communication-efficient model aggregation, thereby achieving much lower communication latency as compared with the OMA schemes \cite{zhu2020broadband, yang2020federated, mohammad2020machine, sery2020analog, zhang2021gradient, fang2022communication}.
Meanwhile, to implement efficient joint learning and communication optimization for AirComp-assisted FL, the convergence analyses and learning-oriented transceiver design are developed in \cite{cao2022optimized, zou2022knowledge, jing2022federated, wang2022interference, wang2022edge, yang2022secondorder} to ensure the model aggregation accuracy while accelerating the model convergence.
The scheduling policy is also developed in \cite{zhu2020broadband, yang2020federated, su2022data}, targeting to restrict the model aggregation error caused by participating devices with varying channel conditions.
Besides, auxiliary equipment like half-duplex relay and RIS are leveraged in \cite{lin2022relayassisted, wang2022federated, liu2021reconfigurable, ni2021federated, hu2022ris, zhao2022performance} to further increase the aggregation accuracy in severe propagation environment, and AirComp-assisted decentralized FL is studied in \cite{ozfatura2020decentralized, shi2021overtheair, xing2021federated} to realize efficient model exchange without the coordination of a central server.
In addition, one-bit quantization based digital AirComp is applied in \cite{zhu2021onebit, adeli2022multicell, sahin2021distributed, sahin2021overtheair, zhao2021broadband} by transplanting the signSGD algorithm \cite{bernstein2018signsgd} to wireless FL system, which effectively relieves the dependence on CSI for transceiver design, which achieves efficient model aggregation at the cost of the model precision.

\subsubsection{Privacy Preservation}

Although the learning tasks in EI networks is conducted distributively without exposing the raw data, the risk of privacy leakage still exists in such scenarios, where the confidential data can be potentially inferred from the publicly shared model/gradient parameters \cite{fredrikson2015model, abadi2016deep, zhu2019deep}.
To quantify the information leakage, \textit{differential privacy} (DP) is a well-established criterion to measure the sensitivity of the statistical change of the dataset with a fresh input \cite{dwork2014algorithmic}.
A certain level of DP can be guaranteed by introducing perturbations into the aggregated model to maintain the disclosed statistics, thereby masking the contribution of arbitrary individual data.
This leads to a trade-off between the privacy and learning performance due to additional artificial local perturbation introduced to transmitted signals for camouflaging private information \cite{wei2020federated}.
Meanwhile, additional power consumption is needed to provide such an artificial noise mask, which may cause insufficient power at devices for subsequent computation and communication.

Fortunately, AirComp allows the individual data to be hidden in the superimposed signal, which prevents the eavesdroppers from accessing a specific user's model information.
Besides, the inherent non-eliminated channel noise can be regarded as a random perturbation conducted on the received signal for masking aggregation information.
Accordingly, the authors in \cite{liu2021privacy} demonstrate that the channel noise in AirComp-assisted FL contributes to DP requirements, which reduces or even completely saves the energy for adding local artificial perturbations.
Besides, the theoretical analyses derived in \cite{seif2020wireless} and \cite{chen2022decentralized} show that the privacy leakage per user scales as $\mathcal{O}(1/\sqrt{K})$ in AirComp-assisted FL, but scales as a constant and does not decay with the number of devices, $K$, in the counterpart OMA schemes. 
Due to the inherent channel noise is uncontrollable during the model aggregation, the adaptive power control is further investigated in \cite{koda2020differentially, liu2021privacy, yang2022differentially} to ensure that the privacy level can be satisfied within different training rounds.
Moreover, the authors in \cite{elgabli2021harnessing} present a scalable FL framework based on uncoded AirComp and \textit{alternating direction method of multipliers} (ADMM) algorithm, which also reveals that the wireless channel interference and perturbations can be harnessed to increase the spectral efficiency while ensuring the privacy requirements.

\subsection{Summary and Challenges}

Broadly speaking, AirComp can be employed for efficient distributed sensing in wireless sensor networks, and applied to facilitate network-wide data/state consensus in wireless control systems. 
Besides, AirComp is capable of enabling efficient model aggregation in distributed learning systems, while preserving the privacy by hiding individual information in the superimposed signal and masking the data with additive channel noise. 
During the course of applications, several challenges also need to be addressed as follows:
\begin{itemize}
	\item \textbf{Customized Performance Metrics:}
	Instead of simply demanding accurate function computation, the intelligent services generally are task-oriented, where the factors such as model convergence rate, DP, and data heterogeneity need to be jointly considered in conjunction with the aggregation accuracy of AirComp.
	Hence, customized performance metrics are needed to enable efficient system design for diverse tasks.
	
	\item \textbf{Flexible Device Participation:}
	During the long-term services, devices may leave early from or arrive late in the service area, which may lead to divergences in data analytic and model training.
	This makes it challenging to devise an adaptive design for AirComp in dynamic systems to increase communication efficiency while guaranteeing system performance.
\end{itemize}


\section{Future Research Directions} \label{sec:future}

With the emergence of new scenarios and technologies, we believe that there are several interesting topics to be explored in the future.
In this section, we present future research directions yet to be further investigated on AirComp.

\subsection{Learning for AirComp}

The existing literature on AirComp mainly adopts convex optimization based methods to implement system design, which generally suffers from a high computational complexity, especially when the design variables are high dimensional.
This ruthlessly creates a gap between the theoretical analysis and the practical demand for real-time signal processing.
Besides, the transmit and receive scalars/matrices are usually coupled in the optimization problem of AirComp, which can be solved by optimizing the transceiver in an alternative manner with multiple iterations until convergence.
However, the obtained solutions are usually sub-optimal in this case.
Meanwhile, a set of real-time input parameters, e.g., instantaneous CSI, need to keep invariant in practice during the entire optimization procedures to ensure that the optimized results are still feasible in the current transmission environment.
The above issues make the convex optimization based methods somewhat inefficient and difficult to be applied in practice.
Hence, learning to optimize has recently been recognized as a promising solution to addressing the resource allocation and transceiver design for future wireless networks \cite{chen2019artificial, shi2022algorithm, wang2022decentralized}.
Specifically, ML techniques are capable of building complicated functional relationships implied in the input and output data, which enable the wireless network to achieve efficient system design by modeling the transmitter, channel, and receiver as one \textit{deep neural network} (DNN).
This raises several open problems that include but not limited to: 1) Robust end-to-end transceiver design for AirComp based on statistical CSI; 2) ML-based AirComp with scalable inference learning models; and 3) Systematic guidance on neural network design for efficient AirComp transceiver design.

\subsection{AirComp for Edge Inference}

ML techniques have been broadly employed to the optimization of wireless communication networks, where the ML solutions are generally obtained from the well-trained DNNs by the inference.
Herein, the inference generally requires the input data to pass through several layers of DNNs to realize a composite function computation.
As the function computation can be achieved during wireless communication via AirComp, it is natural to consider the possibility of shifting part of inference into the air.
Motivated by this, the authors in \cite{sanchez2022airnn} develop an over-the-air convolution method, called AirNN, to shift the convolution operation of \textit{convolutional neural networks} (CNNs) from devices into the ambient environment.
To achieve the desired convolution operation, multiple programmable RISs are deployed to modify the reflecting signals to ensure that the superimposed signal can be combined in a deterministic manner at the FC, thereby resembling the procedure of input data passing through a convolutional layer \cite{sanchez2022airnn}.
This work is a step forward in the use of AirComp to facilitate the ML process at wireless terminals.
Nonetheless, there are still several challenges to be addressed in achieving the above convolution operations, such as the precise synchronization requirement, energy consumption and latency considerations, and the scalability of practical applications.
Moreover, it remains an open issue to explore the feasibility of AirComp to compute different kinds of layers in DNNs, which leads to open problems including: 1) Transceiver design for achieving the inference of various neural network layers; and 2) Analyses of the influence of AirComp-enabled forward/backward layer propagation on inference accuracy.

\subsection{Secure and Resilient AirComp}

In foreseen data-driven intelligent wireless networks, data security and privacy are important issues to be considered beyond the requirement of efficiency and reliability during the data transmission.
By harnessing the intra-cell interference, it is natural for AirComp to provide inherent privacy preservation through hiding individual data in the superimposed signals masked by the receiver noise \cite{zhu2021overtheair}, which effectively prevents eavesdroppers from recovering information of a specific device.
Nevertheless, there is still a risk of leakage of the function computation results, which needs to prevent eavesdroppers from obtaining accurate computation results.
To this end, the authors in \cite{frey2020towards} propose to include a jammer in AirComp system, which ensures that the jamming signal can be canceled at the legitimate receiver but is indistinguishable from white noise at the eavesdropper by exploiting two information theoretical tools, i.e., coding for the compound channel and channel resolvability.
In addition to the initial theoretical analysis and outlook on secure AirComp presented in \cite{frey2020towards}, the
authors in \cite{hu2022secure} propose a transceiver design for secure AirComp to minimize the computation error at the FC with a guaranteed error at the eavesdropper above a threshold.
Besides, adversarial devices may exist to destroy AirComp by transmitting irrelevant signals over the same radio channel, thereby leading to invalid function computation at the FC, making it necessary to exclude adversarial signals from the the received superimposed signal.
The above issues motivate some open problems including: 1) Secure AirComp with untrusted intermediate nodes in hierarchical networks; and 2) Attack-resilient AirComp in the presence of adversarial devices.

\subsection{Integrated Sensing/Communication and AirComp}

Except for the investigation on co-existed services provided by dedicated devices in a cellular network, recently there has been a focus on how to enable the devices to provide integrated multi-functional services.
One of the representative studies is the \textit{integrated sensing and communication} (ISAC) technique \cite{liu2022integrated}, which captures the integration gain to improve the efficiency of spectrum and hardware utilization, and the coordination gain to implement mutual assistance while balancing the dual-functional performance.
This motives the exploration on how to build a multi-functional network by integrating AirComp and other services with multi-modal devices.
Thereupon, the authors in \cite{qi2020integration} and \cite{qi2021integrated} consider a dual-functional task by integrating communication and AirComp, where the signals transmitted by each device consisting of a computation signal for AirComp and a communication signal for information upload.
As channel interference has different impacts on the communication and AirComp, transceiver beamformings are optimized for coordinating the interference to minimize the MSE of AirComp while maximizing the SINR/sum-rate of communication services \cite{qi2020integration, qi2021integrated}.
In addition, the authors in \cite{li2022integrated} integrate radar sensing and AirComp to realize target detection and function computation simultaneously, which demonstrates the effectiveness of integration framework for improving the spectral efficiency.
The study on integrated sensing/communication and AirComp is still in its infancy, which poses several open problems: 1) Transceiver design for FD integrated sensing and AirComp; 2) Analyses on fundamental performance tradeoff of integrated sensing/communication and AirComp; and 3) Waveform design for integrated sensing/communication and AirComp.


\section{Conclusions} \label{sec:conclusion}

A systematic overview and research outlook of AirComp systems were presented in this paper. 
We first introduced the basics of AirComp from the computation design and communication design perspectives, which laid the groundwork for subsequent developments of AirComp.
Then, AirComp over different network architectures were elaborated in terms of single-cell, multi-cell, hierarchical, decentralized, RIS-aided, and UAV-aided networks, and the critical issues were discussed along with the existing works studied in these scenarios.
Subsequently, a literature review on technologies from the information theory and signal processing perspectives were presented, which described the efforts that have been made to analyze and optimize AirComp systems in recent years.
By considering the issues that may be encountered in the practical deployment of AirComp, we further reviewed the related works in terms of efficient channel feedback, robust design, blind data fusion, energy-efficient and sustainable design, and prototype validation.
In addition, we introduced the applications of AirComp in IoT and EI networks, which demonstrated the advantages of AirComp in reducing communication latency, increasing spectral efficiency, and privacy preservation.
Finally, we identified several future research directions to motivate future works of AirComp.

\bibliographystyle{IEEEtran}
\bibliography{refs}

\end{document}